\documentclass[pra,twocolumn]{revtex4}
\usepackage{graphicx}
\usepackage{amsfonts,amsmath,amssymb,float}
\usepackage{bm,dsfont}
\usepackage{color}
\usepackage{bbm}
\usepackage{braket}
\usepackage{longtable}
\usepackage[dvips]{epsfig}
\usepackage{amsmath,amssymb,lscape,float}
\usepackage{hyperref}
\usepackage{listings}

\begin{document}
\title{Extracting spectral properties of small Holstein polarons from a transmon-based 
analog quantum simulator}

\author{Vladimir M. Stojanovi\'c}
\email{vladimir.stojanovic@physik.tu-darmstadt.de}
\affiliation{Institut f\"{u}r Angewandte Physik, Technical University of Darmstadt, 
D-64289 Darmstadt, Germany}

\date{\today}

\begin{abstract}
The Holstein model, which describes purely local coupling of an itinerant excitation (electron, hole, exciton) 
with zero-dimensional (dispersionless) phonons, represents the paradigm for short-range excitation-phonon
interactions. It is demonstrated here how spectral properties of small Holstein polarons -- heavily phonon-dressed 
quasiparticles, formed in the strong-coupling regime of the Holstein model -- can be extracted from an analog quantum 
simulator of this model. This simulator, which is meant to operate in the dispersive regime of circuit 
quantum electrodynamics, has the form of an array of capacitively coupled superconducting transmon qubits and
microwave resonators, the latter being subject to a weak external driving. The magnitude of $XY$-type coupling between 
adjacent qubits in this system can be tuned through an external flux threading the SQUID loops between those
qubits; this translates into an {\em in-situ} flux-tunable hopping amplitude of a fictitious itinerant spinless-fermion 
excitation, allowing one to access all the relevant physical regimes of the Holstein model. By employing the kernel-polynomial 
method, based on expanding dynamical response functions in Chebyshev polynomials of the first kind 
and their recurrence relation, the relevant single-particle momentum-frequency resolved spectral function of this system 
is computed here for a broad range of parameter values. To complement the evaluation of the spectral function, 
it is also explained how -- by making use of the many-body version of the Ramsey interference protocol -- this 
dynamical-response function can be measured in the envisioned analog simulator. 
\end{abstract}
\maketitle

\section{Introduction}
The overarching goal that motivates investigations of analog quantum simulators~\cite{AnalogSimulBOOK:22} is to facilitate 
understanding of static- and dynamic properties of complex, naturally-occurring quantum many-body systems by studying their 
synthetic, artificially engineered counterparts~\cite{Georgescu+:14}. These synthetic systems, which are typically far 
more amenable to manipulation and control than naturally-occurring physical systems, can be realized in physical platforms 
as diverse as cold neutral atoms in optical lattices~\cite{Bloch+Dalibard+Zwerger:08} or tweezers~\cite{MorgadoWhitlock:21}, 
trapped ions~\cite{Bruzewicz+:19}, cold polar molecules~\cite{Gadway+Yan:16}, and low-impedance superconducting (SC) quantum 
circuits (those in which the Josephson energy dominates over the charging energy)~\cite{Wendin:17,Gu+:17,SCqubitReview:19,
SCdevicesPractical:21,SCcircuitsTutorial:21}, to name but a few. In particular, SC analog quantum simulators~\cite{Hohenadler+:12,
Gangat+:13,Kapit:13,Mei+:13,Egger+Wilhelm:13,Paraoanu:14,Stojanovic+:14,LasHeras++:14,Leppakangas+:18,Stojanovic+Salom:19,
StojanovicPRL:20,Nauth+Stojanovic:23} are typically based on arrays of coupled transmon qubits~\cite{Koch+:07} and SC microwave 
resonators, which represent the principal building blocks of circuit-quantum-electrodynamics (circuit-QED) systems~\cite{Wallraff+:04,
Girvin:14,VoolDevoretReview:17
}. 

The time-honored molecular-crystal model due to Holstein~\cite{Holstein:59} describes purely local, density-displacement type coupling 
of an itinerant excitation (electron, hole, exciton) with dispersionless phonons -- i.e., zero-dimensional (Einstein-type) harmonic 
oscillators residing at each site of the underlying lattice. This model describes a smooth crossover from a weakly phonon-dressed 
(quasi-free) excitations to a strongly dressed one ({\em small Holstein polaron}) upon increasing the e-ph coupling strength~\cite{AlexandrovDevreeseBook}. 
In addition, Holstein-type e-ph interaction is known to play an important role for transport properties of certain classes of nonpolar, 
narrow-band electronic materials~\cite{Hannewald:04,Hannewald++:04,StojanovicNonlocal:04,Roesch+:05,Vukmirovic+:10,Stojanovic++:10,
Vukmirovic+:12,Makarov+:15,Shneyder+:20,Shneyder+:21}, often in combination with nonlocal e-ph interaction mechanisms (e.g., 
Peierls-~\cite{Stojanovic+:04,Stojanovic+Vanevic:08,Stojanovic:20} and breathing-mode-type~\cite{Slezak++:06,StojanovicPRA:21} 
e-ph interactions). 

Small polarons are characterized by the following two essential physical features~\cite{AlexandrovDevreeseBook}. Firstly, the 
center of the small-polaron Bloch band is shifted downwards with respect to that of the original bare-excitation band by an amount 
usually referred to as the small-polaron binding energy. Secondly, the width of this (small-polaron) Bloch band is usually much 
smaller than that of its corresponding bare-excitation counterpart, being sometimes even exponentially suppressed with respect to the latter 
bandwidth as a function of the dimensionless e-ph coupling strength. The ground-state properties of Holstein polarons have been extensively 
investigated in the past~\cite{Jeckelmann+White:98,Bonca+:99,Ku+Bonca:02}, for different dimensionalities of the underlying lattice
and using a large variety of analytical and numerical methods. While ground-state properties of small Holstein polarons are 
well-understood by now, various aspects of the Holstein model at finite carrier density~\cite{Zhao+:23,Dee+:23}, as well as
the interplay of Holstein-type e-ph coupling with Hubbard-type electron correlations (the Hubbard-Holstein model)~\cite{Li+:18,
Hebert+:19}, still attract considerable attention. Despite its inherent simplicity, the Holstein model remains the most common 
starting point for discussing polaronic behavior in various classes of electronic materials in which e-ph coupling has short-range 
character~\cite{Kang+:18,Franchini+:21}.

Several analog quantum simulators envisaged to emulate the physics of the Holstein molecular-crystal model~\cite{Holstein:59} 
have been proposed in the past~\cite{Stojanovic+:12,HerreraEtAl,Mei+:13}. They have, however, all been discussed in the context 
of investigating the ground-state properties of this model. At the same time, dynamical and spectral properties of the Holstein 
model have still not received due attention among the workers in the field of analog quantum simulation, despite an increased 
interest in those aspects of the model within the solid-state-physics community~\cite{HuangEtAlNJP:21}. 

In this paper, spectral properties of phonon-dressed excitations governed by the one-dimensional Holstein model are investigated 
within the framework of a SC analog simulator. This simulator is meant to operate in the dispersive (non-resonant) regime of 
circuit QED~\cite{Wallraff+:04
} and allows one to access all the relevant physical regimes of the Holstein model. 
It has the form of an array of capacitively coupled transmon qubits and microwave resonators, with the latter being simultaneously 
subject to a weak external driving. The 
magnitude of the $XY$-type (flip-flop) coupling between adjacent transmon qubits in this system can be tuned by an external 
flux threading SQUID loops between adjacent qubits; this external flux, which allows one to mimic an {\em in-situ} tunable 
hopping amplitude of an itinerant spinless-fermion excitation, represents the main experimental knob in this system. 

To characterize the spectral properties of phonon-dressed excitations resulting from strong Holstein-type (local) e-ph coupling, the 
momentum-frequency resolved spectral function is evaluated here for different choices of parameters characterizing the proposed 
SC analog simulator using the {\em kernel-polynomial method} (KPM)~\cite{Silver+Roeder:94}. This powerful method, based on the expansion 
of the relevant spectral function in Chebyshev polynomials of the first kind~\cite{WeisseEtAlRMP:06} and their recurrence relation, 
was pioneered by Silver and R\"{oder}~\cite{Silver+Roeder:94,Silver+:96,Silver+Roeder:97} and was successfully employed in the past 
for evaluating both zero-~\cite{Alvermann+Fehske:08} and finite-temperature~\cite{Schubert+:05} dynamical response functions of 
various quantum many-body systems~\cite{WeisseEtAlRMP:06}. In recent years, several generalizations of the original KPM have 
also been proposed~\cite{Jiang+:21,Sobczyk+Roggero:22,ZhaoEtAl:23}.

The proposed analysis of spectral properties of Holstein polarons within the framework of a SC analog simulator 
is particularly pertinent because of the availability of a method for the experimental measurements of retarded 
two-time correlation functions based on a generalized, multi-qubit version of the Ramsey interference protocol~\cite{Knap++:13}. 
This method is applicable to all locally-addressable systems, i.e. systems that can be addressed at a single-qubit 
level~\cite{Stojanovic+:14}. In particular, the envisioned SC analog simulator belongs to this class of systems.

The remainder of this paper is organized as follows. In Sec.~\ref{SimulatorLayout} the SC analog simulator 
to be considered in the following sections is introduced, along with its governing Hamiltonian and a 
discussion of typical values of its characteristic parameters. In Sec.~\ref{DeriveHolstein} the derivation 
of the effective Holstein-type Hamiltonian of this system is presented, which is based on the standard 
transformations for the dispersive regime of circuit QED. In Sec.~\ref{SpectFuncEval} the most general properties 
of the momentum-frequency resolved spectral function are briefly reviewed, this being followed by the layout of
the scheme for its measurement in the envisioned analog simulator, and, finally, a short description of the numerical 
strategy for computing this dynamical response function using the KPM. The results obtained 
through the numerical evaluation of this spectral function are presented and discussed in Sec.~\ref{resdiscuss}. 
Finally, the paper is summarized in Sec.~\ref{SumConcl}, with some salient concluding remarks and future outlook. 
Some mathematical details related to the truncation of the (infinite-dimensional) Hilbert space of the 
coupled e-ph system under consideration and its symmetry-adapted basis are relegated to Appendix~\ref{ExactDiag}.
In addition, a brief review of the many-body Ramsey interference protocol is provided in Appendix~\ref{Extract_via_Ramsey},
while the basics of the KPM and its application in evaluating dynamical-response functions are recapitulated 
in Appendix~\ref{KPMbasics}.

\section{Analog quantum simulator} \label{SimulatorLayout}
The principal building blocks of the envisioned simulator (for a schematic illustration, see 
Fig.~\ref{fig:circuit} below) are SC transmon qubits ($Q_n$) with the energy splitting $\hbar\omega_{z}$, 
microwave resonators ($R_{n}$) with the photon frequency $\omega_{c}$ (assumed to be realized in the the 
form of coplanar waveguides~\cite{PozarBook}), and SQUID loops ($J_{n}$), each of which comprises two 
Josephson junctions with the energy $E^{0}_{J}$ ($n=1,\ldots,N$). The pseudospin-$1/2$ degree of freedom 
of the $n$-th qubit will hereafter be represented by the Pauli operators $\bm{\sigma}_n\equiv (\sigma^{x}_{n},
\sigma^{y}_{n},\sigma^{z}_{n})$. At the same time, microwave photons in the resonators, created (annihilated)
by the operators $a_{n}^{\dagger}$ ($a_{n}$), play the role of dispersionless (Einstein-type) phonons. 

The total Hamiltonian of the envisioned simulator can be written in the form 
\begin{equation} \label{eq:HtotDef}
H_{\textrm{tot}}=\sum_n\big(H^{0}_{n} + H^{d}_{n} + H_{n,n+1}^{J} \big) \:,
\end{equation}
where $H^{0}_{n}$ accounts for the $n$-th qubit, its corresponding resonator, and
their mutual (qubit-resonator) always-on coupling, $H^{d}_{n}$ describes external 
microwave driving of the $n$-th resonator, while $H_{n,n+1}^{J}$ describes the 
coupling between qubits $n$ and $n+1$ through the SQUID loop $J_n$. 

\begin{figure}[t!]
\includegraphics[clip,width=8.4cm]{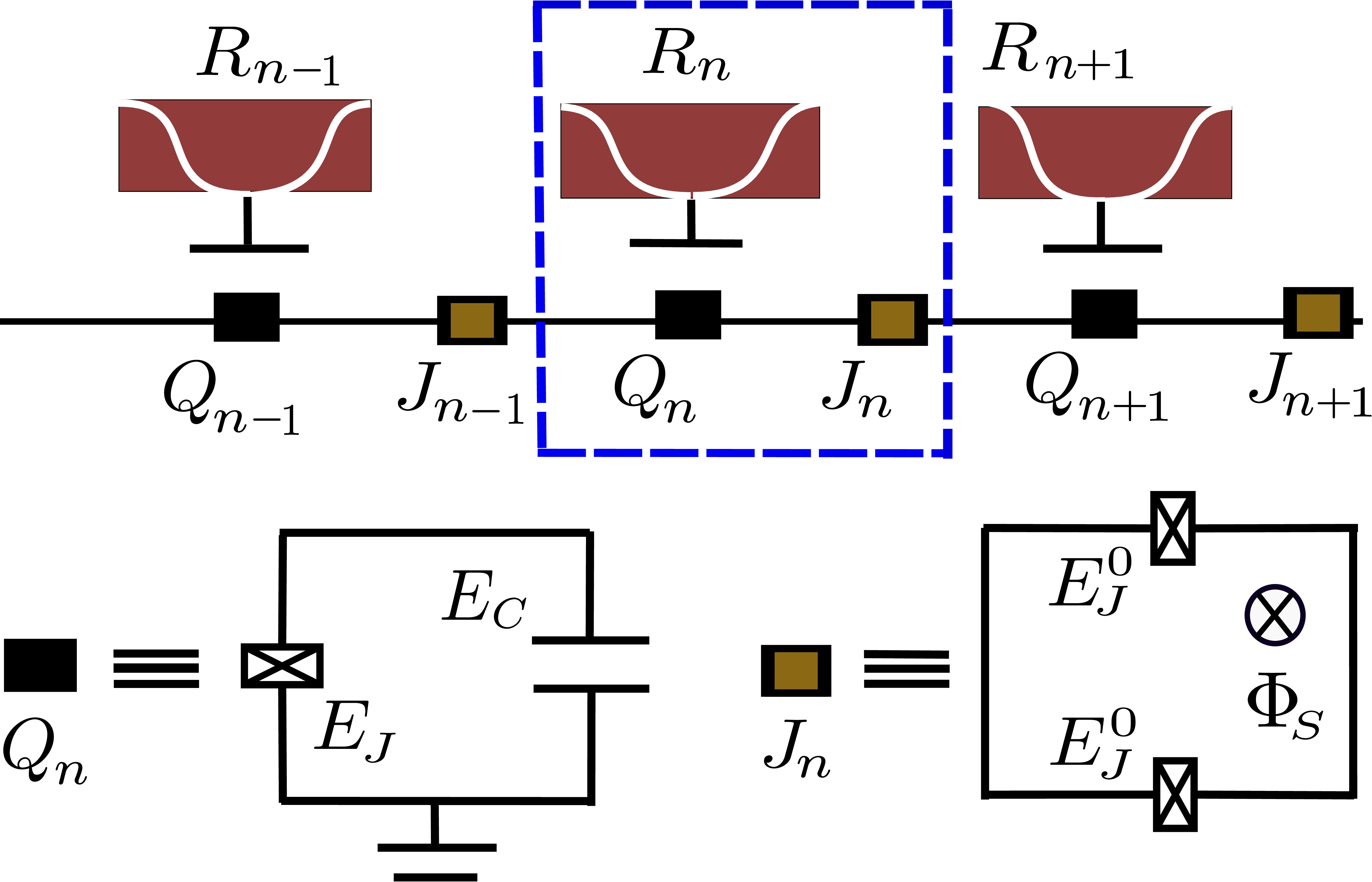}
\caption{\label{fig:circuit}Schematic of the envisioned SC analog simulator, whose
$n$-the repeating unit (indicated by the dashed rectangle) consists of the transmon qubit $Q_{n}$, 
the microwave resonator $R_{n}$, and the SQUID loop $J_{n}$ that comprises two Josephson junctions with 
the energy $E^{0}_{J}$. These SQUID loops are threaded by an external magnetic flux of magnitude 
$\Phi_S$, the main experimental knob in this system. The single-qubit charging- and Josephson 
energies are denoted by $E_{C}$ and $E_{J}$, respectively.}
\end{figure}

Here $H^{0}_{n}$ is given by a Hamiltonian of the Jaynes-Cummings form~\cite{JaynesCummingsBOOK:21}:
\begin{eqnarray}\label{eq:Hn}
H_{n}^{0}=\frac{\hbar\omega_{z}}{2}\:\sigma^{z}_{n} + \hbar\omega_c\:a_{n}^{\dagger}a_{n}
+g\left(a_{n}^{\dagger}\sigma^{-}_{n}+a_{n}\sigma^{+}_{n}\right) \:.
\end{eqnarray}
While the first two terms on the right-hand-side (RHS) of the last equation describe the 
$n$-th qubit and its corresponding resonator, respectively, the final term accounts 
for the capacitive qubit-resonator coupling~\cite{SCqubitReview:19}; in particular, 
$g$ is the qubit-resonator coupling strength. At the same time, the resonators are driven 
by an ac microwave source described by the time-dependent Hamiltonian 
\begin{equation} \label{eq:Hdrive}
H^{d}_{n}=2\xi_d \cos(\omega_d t)(a_{n}+a_{n}^{\dagger}) \:,
\end{equation}
where $\xi_d$ is the driving amplitude and $\omega_d$ its corresponding frequency.

The Hamiltonian $H_{n,n+1}^{J}$, which describes the coupling between adjacent qubits
$n$ and $n+1$, can be written as 
\begin{equation}\label{eq:HJ} 
H_{n,n+1}^{J} = -E_{JS}\cos(\varphi_{n}-\varphi_{n+1})  \:,
\end{equation}
where $\varphi_{n}$ is the gauge-invariant phase of the SC island of the $n$-th qubit 
and $E_{JS}$ the effective Josephson energy of the SQUID loop. 
In the regime of interest for transmon qubits ($E_J/E_C\sim 100$, where $E_J$ and $E_C$
are the Josephson- and charging energies of a single qubit, respectively), the term 
$\cos(\varphi_{n}-\varphi_{n+1})$ is well-approximated by an expansion up to the second order 
in $\varphi_{n}-\varphi_{n+1}$, where this expansion involves $\zeta^{2}_{0}\equiv 
(2E_{C}/E_{J})^{1/2}$ (where $\zeta_{0}\equiv (2E_{C}/E_{J})^{1/4}$ is the quantum 
displacement of the gauge-invariant phase variables~\cite{Girvin:14}) as the relevant 
small parameter (for a typical transmon $\zeta^{2}_{0}\sim\:0.15$); 
higher-order terms in that expansion, i.e. higher powers of the phase difference 
$\varphi_{n}-\varphi_{n+1}$, can safely be neglected owing to their rapidly decaying 
prefactors, which are proportional to higher powers of $\zeta^{2}_{0}$.
In this manner, upon switching to the pseudospin-$1/2$ (qubit) operators $\bm{\sigma}_n$, 
the Hamiltonian in Eq.~\eqref{eq:HJ} can approximately be rewritten in the form (for a 
detailed derivation, see Ref.~\cite{Nauth+Stojanovic:23})
\begin{equation} \label{QubCouplApprox}
H_{n,n+1}^{J} \approx E_{JS}\zeta^{2}_{0}\Big[\frac{\sigma^{z}_{n}+\sigma^{z}_{n+1}}{2} 
-\big(\sigma^{+}_{n}\sigma^{-}_{n+1}+\sigma^{-}_{n}\sigma^{+}_{n+1}\big)\Big]  \:,
\end{equation}
where constant terms -- immaterial for our present purposes -- have been dropped.
Assuming that every Josephson junction in the SQUID loops has the same energy $E^{0}_{J}$ and that 
an external magnetic flux of magnitude $\Phi_{S}$ is threading each loop (cf. Fig.~\ref{fig:circuit}), 
the effective ($\Phi_{S}$-dependent) Josephson energy $E_{JS}\equiv E_{JS}(\Phi_{S})$ of a loop is 
given by~\cite{Makhlin+:01}
\begin{equation}\label{eq:E_J}
E_{JS}=2E^{0}_{J}\:\cos(\pi\Phi_{S}/\Phi_{0}) \:, 
\end{equation}
where $\Phi_{0}\equiv hc/(2e)$ is the magnetic-flux quantum.

It is important to point out that in the derivation of Eq.~\eqref{QubCouplApprox} the terms of the 
form $\sigma_{n}^{+}\sigma_{n+1}^{+}$ and $\sigma_{n}^{-}\sigma_{n+1}^{-}$ have been dropped. 
While this is permitted on the grounds of the rotating-wave approximation (RWA) -- even in the most 
rigorous (multilevel) treatment of transmon qubits -- in the problem under consideration there is an additional argument for neglecting 
such terms. Namely, in the single-excitation problem under consideration (see Sec.~\ref{HolsteinHamParameters} 
below) the relevant part of the total Hilbert space of the system comprises states with a single spinless 
fermion -- which in the pseudospin-$1/2$ (qubit) representation correspond to states with exactly one 
qubit in the logical state $|1\rangle$ (i.e. these states correspond to Hamming-weight-$1$ bit 
strings) -- and the terms of the type $\sigma_n^- \sigma_{n+1}^-$ and $\sigma_n^+ \sigma_{n+1}^+$ yield 
zero when acting on an arbitrary such state.

It is also worthwhile to note that the $\sigma^{z}_{n}$ terms in Eq.~\eqref{QubCouplApprox} are of the 
same form as the single-qubit terms in $H_{n}^{0}$ [cf. \eqref{eq:Hn}]; in other words, this type
of qubit-resonator coupling leads to a shift in the qubit frequency. Therefore, it is pertinent 
to hereafter absorb this shift into the coefficients in front of the single-qubit terms in Eq.~\eqref{eq:Hn}.
The total Hamiltonian of the analog simulator under consideration [cf. Eq.~\eqref{eq:HtotDef}] is 
then given by [cf. Eqs.~\eqref{eq:Hn} - \eqref{QubCouplApprox}]
\begin{equation} \label{eq:Htot}
H_{\textrm{tot}}=\sum_n\big[H^{0}_{n}+H^{d}_{n}-E_{JS}\zeta^{2}_{0}\big
(\sigma^{+}_{n}\sigma^{-}_{n+1}+\sigma^{-}_{n}\sigma^{+}_{n+1}\big)\big] \:.
\end{equation}
The last term on the RHS of this equation, with the prefactor $E_{JS}\zeta_{0}^{2}$, 
describes the $XY$-type (flip-flop) coupling between adjacent transmon qubits.

\section{Effective Holstein-like Hamiltonian} \label{DeriveHolstein}
In the following, the effective Hamiltonian of the envisioned analog simulator in the dispersive 
regime of circuit QED is first derived (Sec.~\ref{EffecHderivation}). This is followed by the 
identification of this effective model with the Holstein Hamiltonian via the Jordan-Wigner 
transformation~\cite{ColemanBOOK} and a discussion of the relevant parameter regimes 
(Sec.~\ref{HolsteinHamParameters}).

\subsection{Derivation of the effective system Hamiltonian} \label{EffecHderivation}
The proposed simulator is meant to operate in the {\em dispersive regime} of circuit QED
, which is defined by the condition that the detuning $\Delta\equiv \omega_c - \omega_z$ between the 
resonator and qubit frequencies is much larger than the qubit-resonator coupling strength $g$, i.e. 
$|\Delta| \gg g$.

The standard practice in the treatment of the Jaynes-Cummings model [cf. Eq.~\eqref{eq:Hn}] and some of 
its generalizations in the dispersive regime, is to first apply a Schrieffer-Wolff-type canonical transformation 
$H_{\textrm{JC}}\rightarrow \widetilde{H}_{\textrm{JC}} \equiv U H_{\textrm{JC}} U^{-1}$~\cite{Klimov+:02}, 
defined by a certain generator $S$ (i.e. $U\equiv e^{S}$), such that the term linear in the light-matter 
coupling completely vanishes. One such generalization is the quantum Rabi model~\cite{Rabi:37,WallsMilburnBook,
Richer+DiVincenzo:16}, which -- unlike the Jaynes-Cummings model itself -- also includes the off-resonant 
qubit-photon interaction processes (i.e. the counter-rotating terms $\sigma^{+}_{n}a_{n}^{\dagger}$ 
and $\sigma^{-}_{n}a_{n}$), so that the resulting qubit-resonator interaction assumes the form 
$g\sigma^{x}_{n}(a_{n}+a_{n}^{\dagger})$ (the so-called transverse coupling term~\cite{Richer+DiVincenzo:16}). 
In particular, in the case of the Jaynes-Cummings model the desired canonical-transformation 
generator is given by~\cite{Boissonneault+:09,Hausinger+Grifoni:10}
\begin{equation}  \label{UnitaryGenerator}
S=\sum_n \frac{g}{\Delta}\:(\sigma^{-}_{n}a_{n}^{\dagger}-\sigma^{+}_{n}a_{n}) \:.
\end{equation}
In the following, with the aim of obtaining an effective Holstein-type Hamiltonian of the envisioned 
analog sumulator, this same canonical transformation will be applied to the total Hamiltonian 
$H_{\textrm{tot}}$ of the system [cf. Eq.~\eqref{eq:Htot}]. This transformation will be carried out 
by making use of the well-known operator identity 
\begin{equation}
 e^{A}B e^{A}=B+[A,B]+\frac{1}{2}\:[A,[A,B]]+\ldots\:,
\end{equation}
where the ellipses on the RHS of the last equation denote higher-order repeated commutators of 
the operators $A$ and $B$. More specifically yet, the Hamiltonian $\widetilde{H}_{\textrm{tot}}\equiv 
UH_{\textrm{tot}} U^{-1}$ will be obtained by keeping only the leading-order terms in the small 
parameter $g/\Delta$.

By carrying out the canonical transformation with the generator given by Eq.~\eqref{UnitaryGenerator}, the 
Jaynes-Cummings contribution $H^{0}_{n}$ to the total system Hamiltonian is transformed into the Hamiltonian 
$\widetilde{H}^{0}_{n}\equiv U H^{0}_{n} U^{-1}$. The latter is given by 
\begin{equation}  \label{tildeHon}
\widetilde{H}^{0}_{n}= \hbar\omega_c\:a_{n}^{\dagger}a_{n} 
+\frac{\hbar}{2}(\omega_z -\chi)\sigma^{z}_{n} - \hbar
\chi\sigma^{z}_{n}a_{n}^{\dagger}a_{n} \:,
\end{equation}
where $\chi \equiv g^{2}/\Delta$ is the Stark shift
. The last term on the RHS of Eq.~\eqref{tildeHon} effectively renders the transition frequency
of the resonator dependent on the state of the qubit and vice versa
. The very existence of this term demonstrates that qubits and resonators are inevitably 
getting entangled through their always-on interaction. 

One can derive $U H^{d}_{n} U^{-1}$ and $U H^{J}_{n,n+1} U^{-1}$ in a similar fashion, 
obtaining as a result the transformed total Hamiltonian $\widetilde{H}_{\textrm{tot}}$.
It is pertinent to analyze this transformed Hamiltonian in the interaction picture --
i.e. the rotating frame -- defined by the Hamiltonian 
\begin{equation}  \label{H_0}
H_{0}= \sum_n \left(\hbar\omega_d\:a_{n}^{\dagger}a_{n}
+\frac{\hbar\tilde{\omega}_z}{2}\:\sigma^{z}_{n}\right)  \:,
\end{equation}
where $\omega_{\Delta}\equiv \omega_c +\chi - \omega_d$ is the modified detuning
of the resonator frequency from that of the external driving and $\tilde{\omega}_z\equiv 
\omega_z - \chi - 2\chi\xi_{d}^2/(\hbar\omega_{\Delta})^2$ the modified qubit frequency. 
To this end, it is useful to recall that in the rotating frame defined by a time-independent 
Hamiltonian $H_{0}$ the (rotating-frame) counterpart $H^{(\textrm{rf})}$ of a general 
Hamiltonian $H$ is given by~\cite{SCqubitReview:19}
\begin{equation}  \label{H_0}
H^{(\textrm{rf})}= \frac{i}{\hbar}\:\partial_t U_{\textrm{rf}}(t)\:U^{\dagger}_{\textrm{rf}}
(t)+ U_{\textrm{rf}}(t) H U^{\dagger}_{\textrm{rf}}(t) \:,
\end{equation}
where $U_{\textrm{rf}}(t)\equiv e^{iH_0 t/\hbar}$ is the Hermitian adjoint of the 
time-evolution operator corresponding to $H_{0}$.

By means of the RWA, i.e. disregarding rapidly-rotating terms 
such as $a_{n}^{\dagger}\sigma^{-}_{n}$, one finds that the counterpart $\widetilde{H}^{(\textrm{rf})}_{\textrm{tot}}$
of the transformed [by the above Schrieffer-Wolff-type canonical transformation; cf. Eq.~\eqref{UnitaryGenerator}]
total Hamiltonian $\widetilde{H}_{\textrm{tot}}$ of the system in the interaction picture (i.e., in the rotating 
frame defined by the Hamiltonian $H_0$) is given by
\begin{equation}  \label{H_r1}
\widetilde{H}^{(\textrm{rf})}_{\textrm{tot}}= \sum_n \left[H^{\textrm{f}}_{n}+H^{J}_{n,n+1} + 
\xi_d (a_{n}^{\dagger}+a_{n})\right]  \:,
\end{equation}
where the Hamiltonian $H^{\textrm{f}}_{n}$ assumes the form 
\begin{equation}  \label{H1n}
H^{\textrm{f}}_{n}=\hbar\omega_{\Delta} a_{n}^{\dagger}a_{n}+\hbar\chi\left(\frac{\xi_d}
{\hbar\omega_{\Delta}}\right)^2\:\sigma^{z}_{n}-\hbar\chi(1+\sigma^{z}_{n})
a_{n}^{\dagger}a_{n}  
\end{equation}
and $H^{J}_{n,n+1}$ is given by Eq.~\eqref{QubCouplApprox}. 

The next step in the derivation of an effective Hamiltonian of the envisioned SC analog simulator is to apply a unitary transformation 
based on the Glauber's displacement operator $D(\beta)\equiv\exp[\sum_n(\beta a^{\dagger}_{n}-\beta^{*}a_{n})]$, 
with $\beta = -\xi_d/(\hbar\omega_{\Delta})$, in order to shift the resonator modes 
according to $a_{n} \rightarrow a_{n}-\xi_d /(\hbar\omega_{\Delta})$ [recall that, in general, $D(\beta) a_n D^{\dagger}(\beta) 
= a_n+\beta$]~\cite{WallsMilburnBook}. The resulting Hamiltonian $D(\beta) \widetilde{H}^{(\textrm{rf})}_{\textrm{tot}} D^{\dagger}(\beta)$ 
[obtained from the one in Eq.~\eqref{H_r1}] assumes the form of an interacting qubit-resonator (i.e. qubit-photon) Hamiltonian
\begin{eqnarray}  \label{H_r}
H_{\textrm{QR}} &=& \sum_n \hbar\omega_{\Delta} \left[ a_{n}^{\dagger}a_{n} 
+ g_{\textrm{H}}\frac{1+\sigma^{z}_{n}}{2}\:(a_{n}+a_{n}^{\dagger}) \right] \nonumber\\
&-& E_{JS}\zeta^{2}_{0} \sum_n\:\big(\sigma^{+}_{n}\sigma^{-}_{n+1}+\sigma^{-}_{n}
\sigma^{+}_{n+1}\big)\:,
\end{eqnarray}
where $g_{\textrm{H}}\omega_{\Delta} = 2\xi_d\chi/(\hbar\omega_{\Delta})$ and in the 
last derivation it was assumed that $\xi_d\gg \hbar\omega_{\Delta}$.

\subsection{Holstein-like effective Hamiltonian and the relevant parameter regime} \label{HolsteinHamParameters}
The final step in deriving an effective Holstein-like Hamiltonian of the system at hand entails switching
from the pseudospin-$1/2$ operators $(\sigma^{x}_{n},\sigma^{y}_{n},\sigma^{z}_{n})$ representing qubits 
to the operators ($c_{n}^{\dagger}, c_{n}$) representing spinless fermions by making use of the Jordan-Wigner 
transformation~\cite{ColemanBOOK}:
\begin{eqnarray} \label{JWtsf}
\sigma_{n}^{z} &=& 2c_{n}^{\dagger}c_{n}-1\:, \nonumber\\
\sigma_{n}^{+} &=& 2c_{n}^{\dagger}\:e^{i\pi\sum_{l<n}c_{l}^{\dagger}c_l}\:.
\end{eqnarray}
On account of the fact that this last transformation maps $\sigma^{+}_{n}\sigma^{-}_{n+1}+\sigma^{-}_{n}
\sigma^{+}_{n+1}$ into the spinless-fermion hopping term $c_{n}^{\dagger}c_{n+1}+\textrm{H.c.}$,
the Hamiltonian of Eq.~\eqref{H_r} can finally be recast in the form 
\begin{eqnarray}  \label{HolsteinHamiltonian}
H &=& \hbar\omega_{\Delta} \sum_n a_{n}^{\dagger}a_{n} -
t_e(\Phi_{S}) \sum_n\big(c_{n}^{\dagger}c_{n+1}+\textrm{H.c.}\big) \nonumber \\
&+& g_{\textrm H}\hbar\omega_{\Delta} \sum_n c_{n}^{\dagger}c_{n}(a_{n}+a_{n}^{\dagger}) \:,
\end{eqnarray}
characteristic of the Holstein model with a single itinerant spinless-fermion excitation
locally coupled to dispersionless phonons (here emulated by photons in the resonators).
Here $t_e(\Phi_{S})\equiv E_{JS}(\Phi_{S})\zeta^{2}_{0}$, explicitly given by [cf. Eq.~\eqref{eq:E_J}]
\begin{equation}\label{eq:t_e}
t_e (\Phi_{S})=2E^{0}_{J}\zeta^{2}_{0}\:\cos(\pi\Phi_{S}/\Phi_{0}) \:, 
\end{equation}
is the {\em in-situ} flux-tunable (nearest-neighbor) hopping amplitude of a 
spinless-fermion excitation, and 
\begin{equation}  \label{g_H}
g_{\textrm{H}} = \frac{2\xi_d\chi}{\hbar\omega^2_{\Delta}}
\end{equation}
the resulting dimensionless Holstein-type (local) e-ph coupling strength; 
the role of the effective phonon frequency is played by $\omega_{\Delta}$.

Before embarking on the discussion as to which regimes of the Holstein model are realizable 
in the analog simulator at hand, it is useful to start with some general considerations. The 
effective e-ph coupling strength $\lambda_{\textrm{H}}$ in a $D$-dimensional Holstein model
with the excitation hopping amplitude $t_{e}$, the phonon frequency $\omega_{\textrm{ph}}$, 
and the dimensionless coupling strength $g_{\textrm{H}}$ is given by the ratio of the 
polaron binding energy in the strong-coupling regime $g_{\textrm{H}}^2\hbar\omega_{\textrm{ph}}$ 
and the average bare-excitation kinetic energy (equal to the half of the bare-excitation 
bandwidth) $2Dt_{e}$, i.e.,~\cite{AlexandrovDevreeseBook}
\begin{equation}  \label{Lambda_Hdef}
\lambda_{\textrm{H}}= g_{\textrm{H}}^{2}
\:\frac{\hbar\omega_{\textrm{ph}}}{2D t_{e}} \:.
\end{equation}
The criterion for the formation of small Holstein polarons is that the conditions 
$\lambda_{\textrm{H}}>1$ and $g_{\textrm{H}}>1$ are simultaneously satisfied~\cite{AlexandrovDevreeseBook}. 
In the adiabatic regime ($\hbar\omega_{\textrm{ph}}/t_{e} <1$), the condition $\lambda_{\textrm{H}}>1$ 
is more difficult to satisfy -- hence being more restrictive -- than $g_{\textrm{H}}>1$.
On the other hand, $g_{\textrm{H}}>1$ becomes the more restrictive of the two conditions 
in the antiadiabatic regime ($\hbar\omega_{\textrm{ph}}/t_{e}>1$). 

Based on the general expression in Eq.~\eqref{Lambda_Hdef}, the effective Holstein e-ph 
coupling strength in the one-dimensional ($D=1$) system at hand (where the role of $\omega_{\textrm{ph}}$
is played by $\omega_{\Delta}$) is given by 
\begin{equation}  \label{LambdaEPH}
\lambda_{\textrm{H}}\equiv g_{\textrm{H}}^{2}
\:\frac{\hbar\omega_{\Delta}}{2t_{e}} \:.
\end{equation}
This effective coupling strength will be varied in the following by exploiting its dependence 
on the hopping amplitude $t_{e}$; importantly, $t_{e}$ itself can be varied by exploiting its 
dependence on the external flux $\Phi_{S}$ [cf. Eq.~\eqref{eq:t_e}]). By combining Eqs.~\ref{g_H} 
and \ref{LambdaEPH}, one obtains the explicit expression for $\lambda_{\textrm{e-ph}}$ as a function 
of the system parameters:
\begin{equation}  \label{LambdaEPHeps}
\lambda_{\textrm{H}}\equiv\frac{2\chi^2}
{\hbar\omega^3_{\Delta} t_{e}}\:\xi_d^{2} \:.
\end{equation}

It is pertinent to consider at this point a choice of parameter values that allows one to realize the 
relevant regimes of the Holstein model within the framework of the envisioned analog simulator. Firstly, 
for the qubit-resonator coupling strength $g$ in a typical circuit-QED setup one has $g/(2\pi\hbar)
\sim 250$\:MHz
. Secondly, the envisioned simulator will be assumed to operate in 
the regime of weak drives, i.e. for $\xi_{d}/g\sim 1$. For instance, $\xi_{d}/(2\pi\hbar)=600$\:MHz 
is one value of the driving amplitude that satisfies this constraint. The detuning $\Delta$ will 
be assumed to have the value $\Delta/(2\pi)=5$\:GHz, which -- along with the aforementioned value of 
$g$ -- clearly satisfies the condition $\Delta\gg g$ that defines the dispersive regime of circuit QED 
[recall Sec.~\ref{EffecHderivation}]; the corresponding value of the Stark shift $\chi=g^2/
\Delta$ is $\chi/(2\pi) = 12.5$\:MHz. Finally, the value $\omega_{\Delta}/(2\pi)=75$\:MHz will be 
used for the parameter $\omega_{\Delta}$. 

By inserting the chosen parameter values into Eq.~\eqref{g_H} above, it is straightforward to obtain 
the value $g_{\textrm H} = 2.667$ for the dimensionless e-ph coupling strength in the system at hand. 
From this last value for $g_{\textrm H}$, which is kept fixed throughout the following discussion,
different values of the effective coupling strength can be obtained from Eq.~\eqref{LambdaEPH} by 
varying the effective hopping amplitude $t_{e}$ (or, equivalently, by changing the adiabaticity ratio $\hbar
\omega_{\Delta}/t_{e}$). This is physically achieved by varying the external flux $\Phi_{S}$ through 
the SQUID loops (cf. Fig.~\ref{fig:circuit}), which determines the value of $t_{e}$ [cf. Eq.~\eqref{eq:t_e}]; 
therefore, the flux $\Phi_{S}$ is the main experimental knob in the system at hand. Importantly, for 
the chosen values of the relevant system parameters one of the two conditions for the existence of 
small Holstein polarons (namely, $g_{\textrm H}>1$) will always be fulfilled, while the other one 
($\lambda_{\textrm H}>1$) will be satisfied for sufficiently large values of the adiabaticity ratio 
$\hbar\omega_{\Delta}/t_{e}$ (more precisely, $\lambda_{\textrm H}>1$ for $\hbar\omega_{\Delta}/t_{e}
\gtrsim 0.28$).

\section{Spectral function: Definition, measurement, and evaluation}   \label{SpectFuncEval}
In the following, after some preliminary considerations of the implications of the discrete translational 
symmetry in the coupled e-ph system at hand (Sec.~\ref{DiscreteTranslInvEPH}), the most important general 
properties of the momentum-frequency resolved spectral function are briefly reviewed (Sec.~\ref{genspectfunc}). 
It is then explained how this dynamical response function can be experimentally extracted in the envisioned 
analog simulator (Sec.~\ref{SpectFuncSCsimul}). Finally, the relevant details of the evaluation of this 
spectral function using the KPM are described (Sec.~\ref{EvalSFwithKPM}).

\subsection{Implications of discrete translational symmetry}  \label{DiscreteTranslInvEPH}
Direct implications of the discrete translational symmetry of the problem at hand are 
discussed in the following, starting with the momentum-space form of the total Hamiltonian 
of the system. Importantly, in what follows all quasimomenta will be expressed in units of 
the inverse lattice spacing, thus the Brillouin zone of the system corresponds to $(-\pi,\pi]$.

The momentum-space form of total Hamiltonian of the system is given by 
$H=H_{0}+H_{\textrm{e-ph}}$, where 
\begin{equation}\label{EephGeneric}
H_{0} = \sum_{k}\:\epsilon_k c_{k}^{\dagger}c_{k} 
+\hbar\omega_{\textrm{ph}}\sum_{q}\:a_{q}^{\dagger}a_{q} \:, 
\end{equation}
is the noninteracting part, with $\epsilon_k\equiv -2t_e \cos k$ being the (one-dimensional) 
tight-binding dispersion of an itinerant spinless-fermion excitation and $\omega_{\textrm{ph}}$ 
the frequency of zero-dimensional phonons. The interacting (e-ph) part $H_{\textrm{e-ph}}$ assumes 
the generic form 
\begin{equation}\label{EephGeneric}
H_{\textrm{e-ph}} = N^{-1/2}\sum_{k,q}\gamma_{\textrm{e-ph}}(k,q)
\:c_{k+q}^{\dagger}c_{k}(a_{-q}^{\dagger}+a_{q}) \:, 
\end{equation}
where $\gamma_{\textrm{e-ph}}(k,q)$ is the e-ph vertex function. In the special case of the 
Holstein model, the corresponding vertex function is completely independent of $k$ and $q$, 
more precisely $\gamma_{\textrm{H}}(k,q)=g_{\textrm{H}}$.

Regardless of the concrete form of the e-ph vertex function, by virtue of the discrete translational 
symmetry of the system, its total Hamiltonian $H$ ought to commute with the total quasimomentum operator 
\begin{equation}
K_{\mathrm{tot}}=\sum_{k}k\:c_{k}^{\dagger}c_{k}
+\sum_{q}q\:a_{q}^{\dagger}a_{q}  \:.
\end{equation}
Because $[H,K_{\mathrm{tot}}]=0$, the (phonon-dressed) Bloch eigenstates of $H$ 
are also the eigenstates of $K_{\mathrm{tot}}$ (instead of being eigenstates of $\sum_{k}k\:c_{k}^{\dagger}c_{k}$,
as would be the case in the absence of e-ph coupling). In other words, the total Hamiltonian $H$
of the system can be diagonalized in sectors of the total e-ph Hilbert space that correspond to the eigensubspaces 
of $K_{\mathrm{tot}}$. For each eigenvalue $k\in (-\pi,\pi]$ of $K_{\mathrm{tot}}$, the Bloch 
eigenstates $|\psi^{(j)}_k\rangle$ of the coupled e-ph system [in the system at hand described by 
the Hamiltonian $H$ in Eq.~\eqref{HolsteinHamiltonian}] form a complete set of states within 
the corresponding sector of the Hilbert space of the coupled e-ph system.

The fact that the coupled e-ph system at hand has a discrete translational symmetry enables one to 
introduce a symmetry-adapted basis [cf. Appendix~\ref{ExactDiag}]. The use of this basis allows one 
to perform the evaluation of the spectral properties more efficiently than in a generic basis. 

\subsection{Momentum-frequency resolved spectral function}  \label{genspectfunc}
The established framework for characterizing excitations in many-body systems is based on the use
of dynamical response functions~\cite{FoersterBOOK}. The latter are defined through Fourier transforms 
of retarded two-time correlation functions, an important example being furnished by the single-particle
retarded Green's function. 

The single-particle retarded Green's function is employed to, e.g., describe the propagation of a single 
electron (or a hole) in solid-state systems. In the problem under consideration, this Green's function 
will be used for the description of an itinerant spinless-fermion excitation coupled with zero-dimensional 
(dispersionless) bosons residing on sites of a one-dimensional lattice.

In the problem at hand, the relevant single-particle retarded Green's function is defined as
\begin{equation}\label{anticommGF}
G_{+}^{\textrm{R}}(k,t)=-\frac{i}{\hbar}\:\theta(t)\langle
\textrm{GS}|\{ c_k^{\dagger}(t),c_k\}|\textrm{GS}\rangle \:,
\end{equation}
where $c_k^{\dagger}(t)\equiv U^{\dagger}_{H}(t)c_k^{\dagger}U_{H}(t)$ is a single-particle creation 
operator in the Heisenberg representation [with $U_{H}(t)$ being the time-evolution operator corresponding to 
the governing Hamiltonian $H$ of the system] and $|\textrm{GS}\rangle$ the ground state of the system;
$\theta(t)$ is the Heaviside function and $\{\ldots,\ldots\}$ stands for an anticommutator of two operators. 

The single-particle retarded Green's function describes the linear response of the system to the addition of 
a single fermion. Its Fourier transform can formally be evaluated provided that a regularization factor $e^{-\eta|t|}$
is included ($\eta\rightarrow 0^{+}$), i.e.,
\begin{equation}
G_{+}^{\textrm{R}}(k,\omega)=\lim_{\eta\rightarrow
0^{+}}\:\int^{\infty}_{-\infty}G_{+}^{\textrm{R}}(k,t)
e^{i\omega t-\eta|t|}dt \:.
\end{equation}
By evaluating the last Fourier transform with the expression for $G_{+}^{\textrm{R}}(k,t)$
from Eq.~\eqref{anticommGF}, one obtains
\begin{eqnarray}
G_{+}^{\textrm{R}}(k,\omega)&=&\langle\textrm{GS}|c_k\:
\frac{1}{\hbar\omega+i0^{+}+E_0-H}
\:c_k^{\dagger}|\textrm{GS}\rangle \nonumber \\
&+& \langle\textrm{GS}|c_k^{\dagger}\:\frac{1}
{\hbar\omega+i0^{+}-E_0+H}\:c_k|\textrm{GS}\rangle\:,
\end{eqnarray}
where $E_0$ is the ground-state energy of the system. 

The momentum-frequency resolved spectral function, the dynamical response function 
of interest in the problem at hand, is given by the imaginary part of $G_{+}^{\textrm{R}}
(k,\omega)$:
\begin{equation} \label{CpectFuncDef}
A_{+}(k,\omega)=-\frac{1}{\pi}\:\textrm{Im}\:G_{+}^{\textrm{R}}(k,\omega)\:.
\end{equation}
By making use of the special case 
\begin{equation} \label{PlemeljFormula}
\frac{1}{x+i0^{+}} = \mathcal{P}\left(\frac{1}{x}\right) 
- i\pi\delta(x) 
\end{equation}
of the Sokhotski-Plemelj theorem, where $x \in \mathbbm{R}$ and $\mathcal{P}$ stands for the 
principal part, along with the fact that the Bloch eigenstates $|\psi^{(j)}_k\rangle$ of the 
total coupled e-ph Hamiltonian form a complete set of states, one can explicitly express the 
spectral function in terms of the eigenstates and eigenvalues of $H$. More precisely, 
$A_{+}(k,\omega)$ is given by
\begin{equation} \label{MomFreqSpectFunc}
A_{+}(k,\omega)=\sum_j\:|\langle\psi^{(j)}_k|c^{\dagger}_k |0\rangle|
^{2}\delta\left(\omega-E_k^{(j)}/\hbar\right) \:,
\end{equation}
where $E_k^{(j)}$ is the energy eigenvalue corresponding to the eigenstate
$|\psi^{(j)}_k\rangle$.

In connection with the relevant single-particle spectral function in the problem 
at hand [cf. Eq.~\eqref{CpectFuncDef}], which corresponds to adding a single fermion
to the vacuum, a remark is in order here. It is important to first note that in the 
solid-state-physics context (i.e. in real electronic materials) the more relevant 
single-particle spectral function is the so-called electron-removal spectral function 
[denoted by $A_{-}(k,\omega)$], as the latter can be experimentally measured using angle-resolved 
photoemission spectroscopy (ARPES)~\cite{DamascelliReview:04}. However, after a removal 
of an itinerant fermionic (e.g. electronic) excitation in a system that only involves dispersionless
(Einstein-type) phonons -- as is the case for the Holstein model investigated here -- the phonon 
degrees of freedom remain frozen and there is no dynamics left in the system. This also explains
why not much attention has heretofore been devoted to the electron-removal spectral function
of the Holstein model; this last spectral function is physically meaningful only for a 
generalized version of the Holstein model that involves phonons with dispersion~\cite{Bonca+Trugman:22}.
To summarize, in the Holstein-polaron problem under consideration, only the spectral
function $A_{+}(k,\omega)$ [cf. Eq.~\eqref{CpectFuncDef}] makes sense physically.

For the sake of completeness, it is worthwhile to mention the intimate
connection between the spectral function of a coupled e-ph system and its
nonequilibrium dynamics following an e-ph interaction quench. To be more
precise, assuming that the initial state of an e-ph system is a bare-excitation
Bloch state $|\Psi_{k}\rangle\equiv c^{\dagger}_k|0\rangle$ with quasimomentum 
$k\in (-\pi,\pi]$, the spectral function $A_{+}(k,\omega)$ is equal to the Fourier 
transform of the matrix element $\langle\psi(t)|c^{\dagger}_k|0\rangle$, where 
$|\psi(t)\rangle$ is the state of the coupled e-ph system at time $t$; this state is given by
a linear combination of Bloch states $|\psi^{(j)}_k\rangle$ of the phonon-dressed 
excitation with the same total quasimomentum $k$. The module squared of this matrix 
element -- i.e. the survival probability of the initial bare-excitation Bloch state 
with quasimomentum $k$ -- is a special case of the quantity that is known as the 
Loschmidt echo. 

In the envisioned analog simulator that consists of $N$ qubits, the bare-excitation Bloch 
state $|\Psi_{k}\rangle$ -- when recast in terms of pseudospin-$1/2$ (qubit) degrees of 
freedom via the Jordan-Wigner transformation [cf. Eq.~\eqref{JWtsf}] -- corresponds to a 
generalized (twisted) $N$-qubit $W$ state~\cite{Haase++:22}; in particular, the state 
$|\Psi_{k=0}\rangle$ corresponds to an ordinary $N$-qubit $W$ state~\cite{StojanovicPRL:20,Zhang++:23}.

\subsection{Measurement of the spectral function in the superconducting analog simulator}  \label{SpectFuncSCsimul}
Having reviewed the general properties of the momentum-frequency resolved spectral
function in Sec.~\ref{genspectfunc}, it is pertinent to discuss at this point some 
relevant aspects in the proposed SC analog simulator.

To begin with, it is worthwhile noting that -- while the anticommutator retarded Green's function, 
defined in Eq.~\eqref{anticommGF}, is the appropriate choice for spinless-fermion excitations 
discussed here -- its commutator counterpart
\begin{equation}\label{commGF}
G_{-}^{\textrm{R}}(k,t)=-\frac{i}{\hbar}\:\theta(t)\langle\textrm{GS}|
[c_k^{\dagger}(t),c_k]|\textrm{GS}\rangle \:,
\end{equation}
has essentially the same physical content in the problem at hand. Namely,
given that $|\textrm{GS}\rangle$ is a vacuum state, one has $c_k^{\dagger}(t)
c_k|\textrm{GS}\rangle=0$, which readily implies that in the single-particle 
problem under consideration the two retarded Green's functions differ only by 
their sign, i.e., $G_{-}^{\textrm{R}}(k,t)=-G_{+}^{\textrm{R}}(k,t)$. Therefore,
provided that there is a way to determine the retarded commutator Green's 
function $G_{-}^{\textrm{R}}(k,t)$ in the proposed analog simulator, its 
anticommutator counterpart $G_{+}^{\textrm{R}}(k,t)$ is also recovered.

Importantly, $G_{-}^{\textrm{R}}(k,t)$ can experimentally be determined in the 
envisioned simulator using a previously proposed scheme based on the many-body version 
of the Ramsey interference protocol~\cite{Knap++:13}; a special case of this scheme,
adapted for systems of the type considered here, is briefly reviewed in Appendix~\ref{Extract_via_Ramsey}.
More precisely, that scheme allows one to extract the real-space retarded (commutator) 
Green's functions 
\begin{equation} \label{RSretardedGreens}
G_{nn'}^{\textrm{R}}(t)\equiv -\frac{i}{\hbar}\:\theta(t)\langle
\textrm{GS}|[c_n^{\dagger}(t),c_{n'}]|\textrm{GS}\rangle \:,
\end{equation}
which in systems with a discrete translational symmetry depend only on $n-n'$,
by measuring a set of retarded two-time correlation functions defined in terms 
of pseudospin-$1/2$ operators (for details, see Appendix~\ref{Extract_via_Ramsey}).
Once $G_{nn'}^{\textrm{R}}(t)$ is obtained in this manner, the retarded momentum-space 
Green's function $G_{-}^{\textrm{R}}(k,t)$ can be determined (for different quasimomenta $k$) 
through a spatial Fourier transformation; for the reasons stated above, this immediately 
also yields $G_{+}^{\textrm{R}}(k,t)$. Finally, once $G_{+}^{\textrm{R}}(k,t)$ is obtained 
at different times $t$, using a numerical Fourier transform to the frequency domain the 
spectral function $A_{+}(k,\omega)$ can be evaluated for a broad range of frequencies 
based on Eq.~\eqref{CpectFuncDef}.

\subsection{Numerical evaluation of the spectral function $A_{+}(k,\omega)$ using the KPM} \label{EvalSFwithKPM}
In this work the KPM (for a brief review of its basic aspects, see Appendix~\ref{KPMbasics}) is employed 
to compute the momentum-frequency resolved spectral function $A_{+}(k,\omega)$ [cf. Eq.~\eqref{CpectFuncDef}]. 
The relevant details of the KPM-based evaluation of this spectral function are described in what follows.

As a preparatory step for the evaluation of $A_{+}(k,\omega)$ using the KPM, the spectrum of the total
effective Hamiltonian $H=H_0+H_\text{e-ph}$ of the e-ph system at hand has to be mapped to the domain 
$[-1,1]$ of Chebyshev polynomials of the first kind $T_r(x)\equiv \cos(r\arccos x)$~\cite{WeisseEtAlRMP:06}. 
This is done by introducing the Hamiltonian
\begin{equation}\label{eq:H_scaled}
\bar{H} = (1-\varepsilon)\frac{2}
{\mathcal{E}_\text{max}-\mathcal{E}_\text{min}}
\left(H - \frac{\mathcal{E}_\text{max}
+\mathcal{E}_\text{min}}{2}\:\mathbbm{1}\right)\:,
\end{equation}
which is shifted and rescaled with respect to $H$. 
Here $\mathcal{E}_\text{max}$ and $\mathcal{E}_\text{min}$ are the largest and smallest 
eigenvalue of $H$, respectively, which can both be obtained using the Lanczos 
algorithm~\cite{CullumWilloughbyBook}. Finally, $\varepsilon$ is an auxiliary parameter that
allows one to avoid stability problems at the boundaries of the spectrum; the value  
$\varepsilon=0.01$ will be used in what follows. 

By inserting $\tilde{H}$ into the expression for the spectral function [cf. Eq.~\eqref{MomFreqSpectFunc}],
one readily obtains
\begin{eqnarray}\label{eq:spec_fct_scaled}
A_{+}(k,\omega) &=& \frac{2\hbar(1-\varepsilon)}
{\mathcal{E}_\text{max}-\mathcal{E}_\text{min}} \sum_j 
\left| \bra{\psi_k^{(j)}} c^\dagger_k \ket{0} \right|^2 
\notag\\&&\times\:\delta\left(\bar{\omega}-\bar{E}^{(j)}_k\right) \:,
\end{eqnarray}
where $\bar{E}^{(j)}_k$ are the eigenvalues of $\bar{H}$ and 
\begin{equation}\label{eq:scaled_freq}
\bar{\omega} = \frac{2(1-\varepsilon)}{\hbar(\mathcal{E}_\text{max}-\mathcal{E}_\text{min})} 
\left(\hbar\omega - \frac{\mathcal{E}_\text{max}+\mathcal{E}_\text{min}}{2}\right) 
\end{equation}
is a shifted and rescaled frequency. Equation~\eqref{eq:spec_fct_scaled} can be recast  
more succinctly as
\begin{equation}\label{eq:spec_fct_fk}
A_{+}(k,\omega) = \frac{2\hbar(1-\varepsilon)}{\mathcal{E}_\text{max}-
\mathcal{E}_\text{min}}\:f_k(\bar{\omega}) \:,
\end{equation}
where $f_k(x)$ is an auxiliary function, defined as
\begin{equation}
f_k(x) \equiv \bra{0}c_k \delta(x-\bar{H}) c^\dagger_k\ket{0} \:.
\end{equation}
In this last expression, $\delta(x-\bar{H})$ can be expanded into a series of 
polynomials $T_r(x)$ and approximated by $f^{(N_{\textrm{C}})}_k(x)$ with a 
large order $N_{\textrm{C}}$ [cf. Eq.~\eqref{ChebyshevAttenExp}].  

Using the general definition of Eq.~\eqref{ChebyshevMoments}, the Chebyshev moments 
corresponding to $\delta(x-\bar{H})$ can be shown to be of the form 
\begin{equation}\label{eq:Cheb_mom_general}
\mu^{(r)}_k = \bra{0}c_k  T_r(\bar{H}) c^\dagger_k\ket{0}  \quad 
(\:r=0,\ldots,N_{\textrm{C}}-1\:)\:.
\end{equation}
In order to efficiently compute $\mu^{(r)}_k$, one makes use of the well-known 
recurrence relation for Chebyshev polynomials of the first kind~\cite{Shima+NakayamaBOOK:10}:
\begin{equation} \label{Reccurence}
T_{r+1}(x) = 2x T_r(x)-T_{r-1}(x) \quad (\:r\in \mathbbm{N}\:)\:.
\end{equation}
In view of this recurrence relation, it is pertinent to define the states $\ket{\alpha^{(0)}_k}\equiv 
c^\dagger_k\ket{0}$ and $\ket{\alpha^{(1)}_k} \equiv \bar{H}\ket{\alpha^{(0)}_k}$, as well as
\begin{equation}\label{eq:alpha_states}
\ket{\alpha^{(r+1)}_k} \equiv 2\bar{H}\ket{\alpha^{(r)}_k} - \ket{\alpha^{(r-1)}_k} \:,
\end{equation}
for $r=2,\ldots,N_{\textrm{C}}-1$.

It is worthwhile to note that the states in Eq.~\eqref{eq:alpha_states} naturally occur when applying 
the above recurrence relation to Eq.~\eqref{eq:Cheb_mom_general}. Furthermore, their use renders the 
evaluation of the Chebyshev moments by iteration rather straightforward. More precisely, starting from 
$\mu^{(0)}_k=1$ and $\mu^{(1)}_k=\langle\alpha^{(1)}_k|\alpha^{(0)}_k\rangle$, one obtains the following 
expressions:
\begin{eqnarray} 
\mu^{(2r)}_k &=& 2\langle\alpha^{(r)}_k |\alpha^{(r)}_k\rangle - \mu^{(0)}_k \:,\notag\\
\mu^{(2r+1)}_k &=& 2\langle\alpha^{(r+1)}_k|\alpha^{(r)}_k\rangle - \mu^{(1)}_k \:. \label{RecMoments}
\end{eqnarray}
While already the form of Eq.~\eqref{eq:alpha_states} indicates that the matrix-vector multiplication (MVM)
is the essential operation for the implementation of the KPM~\cite{WeisseEtAlRMP:06}, the last equation 
implies that in the problem at hand one can compute two moments from every new state $|\alpha^{(r)}_k\rangle$. 
In other words, two new moments can be obtained by carying out a single MVM. 

The described computational scheme allows a resource-efficient evaluation of the spectral function, given 
that -- being subject to Eq.~\eqref{eq:alpha_states} -- only three states $\ket{\alpha^{(r)}_k}$ have to be 
stored at each step of the evaluation. This is in stark contrast with the Lanczos recursion methods~\cite{CullumWilloughbyBook}, 
where a similar iteration has to be performed, but with a simultaneous preservation of orthogonality at 
each step. Because of that, the KPM is faster and less memory-consuming than the Lanczos methods. Another
important advantage of the KPM compared with the latter methods is that it avoids accumulation of numerical
roundoff errors even in cases where it entails large numbers of MVMs~\cite{Silver+:96}. 

The final step in the evaluation of $f^{(N_{\textrm{C}})}_k(x)$ entails -- in the interest of achieving 
a high numerical efficiency -- a special choice of values for the rescaled frequency, namely
\begin{equation}
\bar{\omega}_j=\cos\left[ \frac{\pi}{N_{\textrm{C}}}\left(j+\frac{1}{2}\right)\right] 
\quad (\:j=0,\ldots,N_{\textrm{C}}-1\:) \:.
\end{equation}
By inserting these last special frequency values into Eq.~\eqref{ChebyshevAttenExp} 
and making use of the identity $T_r(\cos y)=\cos(ry)$ for $y\in[-1,1]$, one arrives
the following result for $f^{(N_{\textrm{C}})}_k(\bar{\omega}_j)$:
\begin{eqnarray} \label{ExprAuxFunc}
&&f^{(N_{\textrm{C}})}_k(\bar{\omega}_j) =\frac{1}{\displaystyle \pi\sin\Big[\frac{\pi}{N_{\textrm{C}}}
\big(j+\frac{1}{2}\big)\Big]} \Big\lbrace \mu^{(0)}_k g_0 \notag\\
&+& 2\sum_{r=1}^{N_{\textrm{C}}-1}\mu^{(r)}_k g_r \cos\left[ r\frac{\pi}{N_{\textrm{C}}}
\left(j+\frac{1}{2}\right)\right]\Big\rbrace \:.
\end{eqnarray}
It is worthwhile noting that this last expression amounts to a discrete Fourier transformation,
which -- as usual -- can be executed with a relatively modest computational effort [namely, with 
$\mathcal{O}(N_{\textrm{C}}\log_2 N_{\textrm{C}})$ operations] using the fast-Fourier-transform (FFT) 
algorithm~\cite{NRcBook}.

With $f^{(N_{\textrm{C}})}_k(\bar{\omega}_j)$ evaluated in the aforementioned fashion, the final 
step in computing the spectral function $A_{+}(k,\omega)$ entails inserting the last result for 
$f^{(N_{\textrm{C}})}_k(\bar{\omega}_j)$ [cf. Eq.~\eqref{ExprAuxFunc}] into 
\begin{equation}  \label{ExprSpectFuncKPM}
A_{+}(k,\omega_j)\approx \frac{2\hbar(1-\varepsilon)}{\mathcal{E}_\text{max}-
\mathcal{E}_\text{min}}\: f^{(N_{\textrm{C}})}_k(\bar{\omega}_j)\:,
\end{equation}
an approximated form of Eq.~\eqref{eq:spec_fct_fk}, where 
\begin{equation}
\bar{\omega}_j = \frac{2\hbar(1-\varepsilon)} {\mathcal{E}_\text{max}
-\mathcal{E}_\text{min}}\:\omega_j - (1-\varepsilon)\:\frac{\mathcal{E}_\text{max}
+\mathcal{E}_\text{min}}{\mathcal{E}_\text{max}-\mathcal{E}_\text{min}}
\end{equation} 
are the shifted and rescaled energy values [cf. Eq.~\eqref{eq:scaled_freq}]. Importantly, while there are
different possible choices of attentuation factors $g_0,\ldots,g_N$ in Eq.~\eqref{ExprAuxFunc} -- hence 
also in Eq.~\eqref{ExprSpectFuncKPM} -- it was demonstrated that those originating from the Jackson 
kernel [cf. Eq.~\eqref{JacksonAttenFactors}] are the optimal choice for evaluating spectral functions 
using the KPM~\cite{Silver+:96}.

It is important to stress that the achievable energy resolution in evaluating spectral functions 
of the type discussed here using the KPM is inversely proportional to the number of Chebyshev moments
used. For a chosen energy resolution, the memory required for Hamiltonians represented by sparse  
matrices can scale linearly in the number of states~\cite{Silver+:96}.

\section{Results and Discussion}     \label{resdiscuss}
In what follows, the results for the momentum-frequency resolved spectral function [cf. Eq.~\eqref{MomFreqSpectFunc}] 
inherent to the Holstein model, obtained using the KPM (for the basic aspects of this method, see Appendix~\ref{KPMbasics}) 
are presented. All the results to be discussed in the following correspond to a system with $N=10$ sites and up to
$N^{\textrm{max}}_{\textrm{ph}}=18$ phonons in the truncated phonon Hilbert space; the dimension of this space is 
$D_{\textrm{ph}}= 43,758$ (a controlled truncation of the Hilbert space of a coupled e-ph system is discussed in
Appendix~\ref{ExactDiag}). To achieve a sufficiently good spectral resolution, as many as $80,000$ Chebyshev moments 
were evaluated and used in the expansion of the spectral function $A_{+}(k,\omega)$ (cf. Sec.~\ref{EvalSFwithKPM}).

The single-particle spectral function is evaluated in the following for five different values of the 
adiabaticity ratio $\hbar\omega_{\Delta}/t_{e}$ (cf. Sec.~\ref{HolsteinHamParameters}), namely $0.125, 0.25,
0.5, 1, 2$. For $\hbar\omega_{\Delta}/t_{e}=0.125$ and $0.25$ the corresponding effective coupling strengths
$\lambda_{\textrm{H}}$ are $0.444$ and $0.889$, respectively; in those cases $\lambda_{\textrm{H}}<1$, thus 
the second condition for the formation of small Holstein polarons is not fulfilled. For the remaining three 
values of $\hbar\omega_{\Delta}/t_{e}$, the system at hand is in the small Holstein-polaron regime 
($g_{\textrm{H}}>1$, $\lambda_{\textrm{H}}>1$).

The coupled e-ph system at hand is treated in what follows under the assumption of periodic boundary conditions (PBC).
Generally speaking, in a system defined on a discrete lattice with $N$ sites, the PBC imply that there are exactly $N$ 
permissible quasimomenta in the Brillouin zone; they are given by $2\pi n/N$, 
where $n=-N/2+1,\ldots, N/2$ ($N$ is assumed to be even). In the system under consideration, with $N=10$,
those quasimomenta are $\{-4\pi/5,-3\pi/5,-2\pi/5,-\pi/5, 0,\:\pi/5,\:2\pi/5,\:3\pi/5,\:4\pi/5,\:\pi\}$.
For each choice of the relevant parameters, the frequency dependence of the spectral function is presented 
in what follows for six different quasimomenta in the positive half of the 1D Brillouin zone of the system
(namely, for $k=0,\:\pi/5,\:2\pi/5,\:3\pi/5,\:4\pi/5,\:\pi$), consistent with the PBC.

\begin{figure}[b!]
\includegraphics[clip,width=0.975\linewidth]{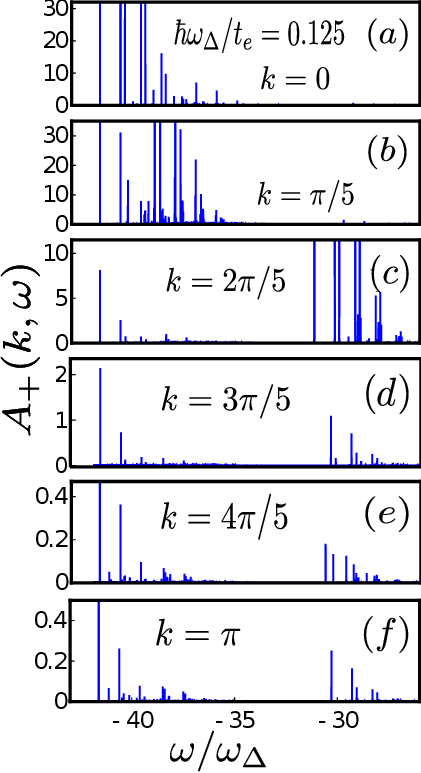}
\caption{\label{fig:SF1} Momentum-frequency resolved spectral function, evaluated for six different quasimomenta $k$
in the positive half of the Brillouin zone and a range of frequencies. The chosen parameter values are $\omega_{\Delta} 
/ (2\pi) = 75$\:MHz, $t_e/(2\pi\hbar) = 600$\:MHz (i.e. the adiabaticity ratio is $0.125$), 
$g/(2\pi\hbar)= 250$\:MHz, $\Delta/(2\pi)=5$\:GHz, and $\xi_d / (2\pi\hbar) = 600$\:MHz. 
The corresponding effective e-ph coupling strength is $\lambda_{\textrm{H}}=0.444$.}
\end{figure}

The numerical evaluation of the single-particle spectral function using the KPM was performed on a $64$-core, 
$7$\:GHz AMD Ryzen Threadripper PRO 5995WX workstation, with the main memory of $527$\:GB; the 
computational runs required to obtain all the results presented in this section took in total about 
$46$ hours. Those results for the spectral function are depicted in Figs.~\ref{fig:SF1} - \ref{fig:SF5} below. 

\begin{figure}[t!]
\includegraphics[clip,width=0.975\linewidth]{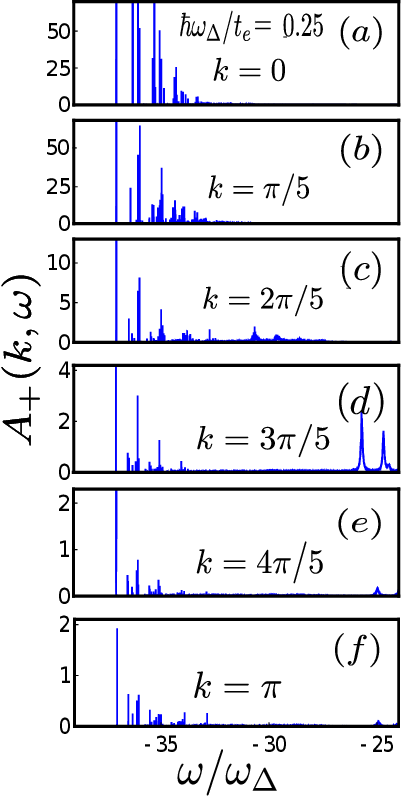}
\caption{\label{fig:SF2} Momentum-frequency resolved spectral function, evaluated for six different quasimomenta $k$
in the positive half of the Brillouin zone and a range of frequencies. The chosen parameter values are $\omega_{\Delta} 
/ (2\pi) = 75$\:MHz, $t_e/(2\pi\hbar) = 300$\:MHz (i.e. the adiabaticity ratio is $0.25$), $g/(2\pi\hbar)= 250$\:MHz, 
$\Delta/(2\pi)=5$\:GHz, and $\xi_d / (2\pi\hbar) = 600$\:MHz. The corresponding effective e-ph coupling strength is 
$\lambda_{\textrm{H}}=0.889$.}
\end{figure}

\begin{figure}[t!]
\includegraphics[clip,width=0.985\linewidth]{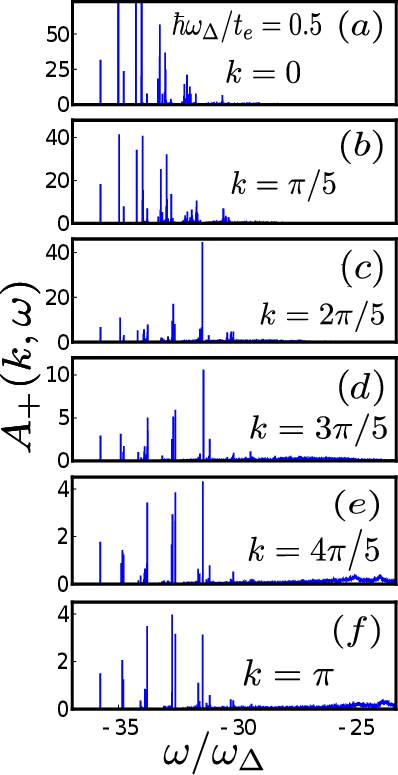}
\caption{\label{fig:SF3} Momentum-frequency resolved spectral function, evaluated for six different quasimomenta $k$
in the positive half of the Brillouin zone and a range of frequencies. The chosen parameter values are $\omega_{\Delta} 
/ (2\pi) = 75$\:MHz, $t_e/(2\pi\hbar) = 150$\:MHz (i.e. adiabaticity ratio is $0.5$), $g/(2\pi\hbar)= 250$\:MHz, 
$\Delta/(2\pi)=5$\:GHz, and $\xi_d/ (2\pi\hbar) = 600$\:MHz. The corresponding effective e-ph coupling strength 
is $\lambda_{\textrm{H}}=1.778$.}
\end{figure}

To make the interpretation of the obtained results more straightforward, it is pertinent to reiterate some of the generic 
properties of the energy spectra of models describing a short-range coupling of an itinerant excitation with Einstein-type 
(zero-dimensional) phonons; the Holstein model, considered in this paper, represents the extreme realization of such models 
in that the Holstein-type e-ph coupling is completely local in real space. In the strong-coupling regimes of such models, 
heavily-dressed excitations (small polarons) are formed. Irrespective of the concrete form of the e-ph coupling in such a 
model, the center of the small-polaron Bloch band is shifted by an energy $E_b$ (the small-polaron binding energy) below 
that of a bare-excitation band.

For each fixed quasimentum $k$, the eigenstates of a coupled e-ph problem that contribute to the spectral function 
$A_{+}(k,\omega)$ [cf. Eq.~\eqref{MomFreqSpectFunc}] include the discrete states (i.e. those belonging to coherent 
Bloch bands of a phonon-dressed excitation) and their corresponding continua. In particular, the energetic width
of each of those continua equals the width of of the respective polaron Bloch band. Importantly, the one-phonon 
continuum corresponds to the inelastic-scattering threshold, i.e. the minimal energy that a phonon-dressed excitation 
must have to be capable of emitting a phonon. The one-phonon continuum sets in at the single-phonon energy 
$\hbar\omega_{\textrm{ph}}$ (where $\omega_{\textrm{ph}}$ is the relevant phonon frequency) above the ground-state 
energy of the coupled e-ph system~\cite{Engelsberg+Schrieffer:63} [recall that in the system under consideration 
the role of the effective phonon freqency is played by $\omega_{\Delta}$ (cf. Sec.~\ref{EffecHderivation})]. 
In the weak e-ph coupling regime of the Holstein- and similar models, a coupled e-ph system only has one discrete 
Bloch state $|\psi^{(j=0)}_k\rangle$ at quasimomentum $k$ -- which in the $k=0$ case represents the ground state 
of the coupled e-ph system -- and its corresponding continuum of states corresponds to a phonon-dressed excitation with 
quasimomentum $k-q$ and an unbound phonon with quasimomentum $q$. Because such a state exists for all possible 
phonon momenta $q$ in the Brillouin zone, in the presence of dispersionless phonons the energetic width of this
continuum is equal to the width of the dressed-excitation Bloch band. 

In the weak-coupling regime, the spectrum of a phonon-dressed excitation is virtually unaffected by the presence of e-ph 
coupling for energies below the phonon-emission threshold. Consequently, in the weak-coupling adiabatic regime ($\hbar
\omega_{\textrm{ph}}<t_e$), the phonon-renormalised dispersion of a spinless-fermion excitation mimics the tight-binding 
cosine dispersion of a bare excitation up to a certain quasimomentum $k_{et}$ that corresponds to the energy $\hbar
\omega_{\textrm{ph}}$ from the bottom of the band. For quasimomenta above $k_{et}$, spinless-fermion- and phonon
states start to hybridise, which leads to the band-flattening phenomenon~\cite{Fehske+:06}.

\begin{figure}[t!]
\includegraphics[clip,width=0.95\linewidth]{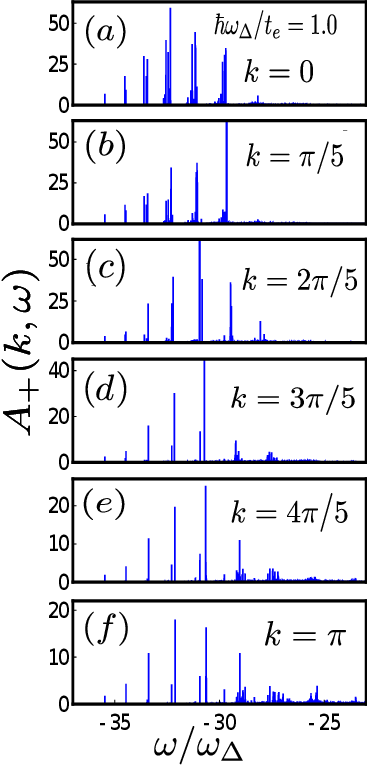}
\caption{\label{fig:SF4} Momentum-frequency resolved spectral function, evaluated for six different quasimomenta $k$ in 
the positive half of the Brillouin zone and a range of frequencies. The chosen parameter values are $\omega_{\Delta} / 
(2\pi) = 75$\:MHz, $t_e/(2\pi\hbar) = 75$\:MHz (i.e. the adiabaticity ratio is $1.0$), $g/(2\pi\hbar)= 250$\:MHz, 
$\Delta/(2\pi)=5$\:GHz, and $\xi_d / (2\pi\hbar)= 600$\:MHz. The corresponding effective e-ph coupling strength is 
$\lambda_{\textrm{H}}=3.555$.}
\end{figure}

\begin{figure}[t!]
\includegraphics[clip,width=0.985\linewidth]{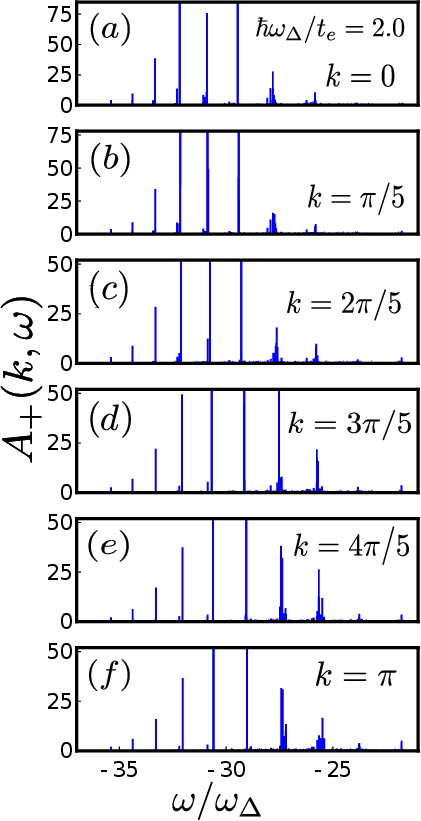}
\caption{\label{fig:SF5} Momentum-frequency resolved spectral function, evaluated for six different quasimomenta $k$ in 
the positive half of the Brillouin zone and a range of frequencies. The chosen parameter values are $\omega_{\Delta} / 
(2\pi) = 75$\:MHz, $t_e/(2\pi\hbar) = 37.5$\:MHz (i.e. the adiabaticity ratio is $2.0$), $g/(2\pi\hbar)= 250$\:MHz, 
$\Delta/(2\pi)=5$\:GHz, and $\xi_d /(2\pi\hbar)= 600$\:MHz. The corresponding effective e-ph coupling strength is 
$\lambda_{\textrm{H}}=7.111$.}
\end{figure}

For larger e-ph coupling strengths, additional coherent polaron bands [i.e., additional discrete (split off from 
the continuum) states at each quasimomentum $k$ in the Brillouin zone of the relevant system] begin to emerge. 
The first such excited dressed-excitation (in the extreme case small-polaron-) state at quasimomentum $k$ 
pertains to a phonon-dressed excitation bound with an additional phonon, such that the total quasimomentum $k$ 
is distributed between the phonon-dressed excitation (quasimomentum $k-q$) and the additional phonon (quasimomentum 
$q$). As e-ph coupling strength continues to increase, another (second) excited state, representing
a bound state of a polaron and two additional phonons (with the same total quasimomentum $k$) sets in. All
those discrete states, together with their respective continua -- separated from them by the energy $\hbar\omega_{\textrm{ph}}$
of a single phonon -- provide additional contributions to the spectral function at the quasimentum $k$~\cite{Fehske+:06}.

In accordance with general characteristics of the energy spectra of (short-range) coupled e-ph Hamiltonians, for increasing 
values of the adiabaticity ratio $\hbar\omega_{\Delta}/t_e$, which in the system at hand translate into larger values of the 
effective e-ph coupling strength [cf. Eq.~\eqref{LambdaEPH}], one can notice an increasing number of discrete peaks in the 
single-particle spectral function. In particular, for the Holstein model it is known that at the onset of the strong-cupling 
regime there are three such peaks. In the system under consideration, from Figs.~\ref{fig:SF4} and \ref{fig:SF5} it can be inferred that 
for the largest effective coupling strength considered ($\lambda_{\textrm{H}}=3.555$ and $\lambda_{\textrm{H}}=7.111$, respectively,
which both belong to the deep strong-coupling regime) there are up to six such peaks, accompanied by their respective continua.

\section{Summary and Conclusions} \label{SumConcl}
In this paper it was demonstrated how spectral properties of small Holstein polarons can be 
investigated using an analog superconducting quantum simulator that consists of capacitively-coupled 
transmon qubits and microwave resonators, the latter being subject to a weak external driving field. 
It was shown here that the envisioned analog simulator, operating in the dispersive regime of circuit 
quantum electrodynamics, allows one to access all the relevant physical regimes of the 
Holstein model. Using the kernel-polynomial method the relevant single-particle momentum-frequency resolved 
spectral function of this system was computed here for a broad range of values for its characteristic parameters. 
Finally, to make contact with anticipated experimental realizations, it was also demonstrated that -- by 
employing the many-body version of the Ramsey interference protocol -- this important 
dynamical-response function can be extracted experimentally in the proposed analog simulator. 

The present study complements investigations of electron-phonon coupling effects in solid-state systems 
in two important respects. Firstly, it allows a nearly-ideal physical realization of the pristine Holstein model 
with purely dispersionless (Einstein-type) phonons, the latter being an idealization for optical phonons in solid-state 
materials. Secondly, it presents results for a specific type of single-particle spectral functions (namely, 
the one corresponding to the addition of a single itinerant excitation to the vacuum state) that cannot be 
directly measured using conventional methods of experimental solid-state physics (such as ARPES), proposing 
also a scheme for the measurement thereof in locally-addressable analog simulators. 

The scheme proposed in this paper can be generalized to other physical realizations of the Holstein model,
i.e. to its analog simulators based on different physical platforms, for example with trapped ions in arrays 
of microtraps. Moreover, recent developments in the field of superconducting qubits~\cite{Wendin:17,Gu+:17,
SCqubitReview:19,SCdevicesPractical:21,SCcircuitsTutorial:21} should make it possible to engineer qubit arrays of 
more nontrivial geometry, that could mimic the geometries inherent to complex electronic materials~\cite{Vanevic+:09}.
Finally, analog simulators that display similar functionalities as transmon-based ones can be engineered
using other types of superconducting qubits. For example, models akin to the Holstein model~\cite{Roosz+Held:22} 
can also be emulated with arrays of flux qubits~\cite{Volkov+Fistul:14,Fistul+:22} and their spectral properties 
investigated based on the general strategy that was laid out here. Experimental realizations of the envisioned 
superconducting analog simulator and the proposed method for extracting the spectral properties of small 
Holstein polarons is keenly anticipated.

\begin{acknowledgments}
The author acknowledges useful discussions on the numerical implementation of the kernel polynomial method
with J. K. Nauth. This research was supported by the Deutsche Forschungsgemeinschaft (DFG) -- SFB 1119 -- 236615297.
\end{acknowledgments}

\appendix 

\section{Truncated Hilbert space and its symmetry-adapted basis} \label{ExactDiag}
The infinite-dimensional character of the phonon (or, more generally, boson) Hilbert spaces
requires one to carry out a controlled truncation of the Hilbert space of the coupled e-ph 
system under consideration. In the following, the essential details of this Hilbert-space
truncation are discussed, along with the introduction of the basis adapted to the discrete
translational symmetry of the system at hand (the symmetry-adapted basis). 

The Hilbert space of the coupled e-ph system under consideration, defined on a lattice 
with $N$ sites, is spanned by states $|n\rangle_e\otimes|\mathbf{m}\rangle_\text{ph}$, 
where $|n\rangle_e\equiv c_{n}^{\dagger}|0\rangle_e$ corresponds to the excitation at 
the site $n$ and 
\begin{equation}\label{mphvect}
|\mathbf{m}\rangle_\text{ph} = \prod_{n=1}^{N\otimes}
\frac{(b_n^\dagger)^{m_n}}{\sqrt{m_n!}}\:|0\rangle_\text{ph}\:,
\end{equation}
with $\mathbf{m}\equiv (m_1,\ldots,m_N)$ being the set of phonon occupation numbers 
at different lattice sites. 

In view of the infinite-dimensional phonon Hilbert spaces, one ought to restrict oneself 
to the truncated phonon Hilbert space that consists of states whose corresponding total 
phonon number $m=\sum_{n=1}^N m_n$ is not larger than $M$, where $0\le m_n \le m$. Consequently, 
the total Hilbert space of the coupled e-ph system has the dimension $D = D_\text{e} \times 
D_\text{ph}$, where for a lattice with $N$ sites $D_\text{e} = N$ and $D_\text{ph}=(M+N)!/(M!N!)$.

It is important to stress that the choice of the maximal total number $M$ of phonons, which 
determines the total dimension of the Hilbert space of a coupled e-ph system, is governed by
the required numerical accuracy in evaluating the desired physical observables (e.g. the ground-state
energy, expected total phonon number in the ground state, quasiparticle spectral weight, etc.). 
The aforementoned controlled truncation of the Hilbert space of a coupled e-ph system is carried 
out by gradually increasing the number of sites $N$, and subsequent increase in the total number
of phonons $M$, up to the point where their further increase does not lead to an appreciable change 
(in accordance with the accepted error margin) in the obtained results for the physical observables 
of interest. 

The dimension of the matrix-diagonalization problem for the total Hamiltonian of the system 
can further be reduced by taking advantage of the discrete translational symmetry of the 
system, mathematically expressed as the commutation $[H,K_{\mathrm{tot}}]=0$
of operators $H$ and $K_{\mathrm{tot}}$. This allows one to diagonalize 
$H$ in Hilbert-space sectors corresponding to the eigensubspaces 
of $K_{\mathrm{tot}}$. The dimension of each of those $K$-sectors of the total Hilbert 
space is equal to the dimension of the truncated phonon Hilbert space (i.e., 
$D_{K}=D_{\textrm{ph}}$). Therefore, it is pertinent to utilize the symmetry-adapted basis 
\begin{equation}\label{symmbasis}
|K,\mathbf{m}\rangle = N^{-1/2} \sum_{n=1}^N e^{iKn}\,\mathcal{T}_{n-1}(|1\rangle_\text{e} 
\otimes |\mathbf{m}\rangle_\text{ph}) \:,
\end{equation}
where $\mathcal{T}_{n}$ ($n=0,1,\ldots, N-1$) are the (discrete) translation operators, whose
action complies with the PBC. Equation~\eqref{symmbasis} can be rewritten in the form
\begin{equation}\label{symmbasalter}
|K,\mathbf{m}\rangle = N^{-1/2} \sum_{n=1}^N e^{iKn}\, |n\rangle_\text{e} 
\otimes \mathcal{T}^{\textrm{ph}}_{n-1}|\mathbf{m}\rangle_\text{ph} \:,
\end{equation}
with the discrete-translation operators $\mathcal{T}^{\textrm{ph}}_{n-1}$ acting on
the phonon Hilbert space. 

In particular, if $|\mathbf{m}\rangle_\text{ph}$ is given by a set of occupation numbers 
\begin{equation}\label{mphvectors}
|\mathbf{m}\rangle_\text{ph} = |m_1,m_2,\ldots,m_{N}\rangle_\text{ph} \:,
\end{equation}
then the $l$-th occupation number corresponding to the state 
$|\mathcal{T}^{\textrm{ph}}_{n-1}\mathbf{m}\rangle$ is given by 
$m_{s(l,n-1)}$, where $s(l,n)$ is defined as
\begin{equation}\label{findef}
s(l,n)\equiv\begin{cases} 
N-n+l\:, \:\text{for $l\le n$} \\
l-n\:, \: \text{for $l>n$}
\end{cases}\:.
\end{equation}

\section{Measurement of $G_{nn'}^{\textrm{R}}(t)$ via many-body Ramsey protocol} \label{Extract_via_Ramsey}
In the following, we recapitulate essential elements of the method for the experimental measurement 
of retarded Green's functions using the many-body (multiqubit) version of the Ramsey interference 
protocol~\cite{Knap++:13}, which in Ref.~\cite{Stojanovic+:14} was adapted to systems of the type 
discussed in the present work.

The real-space retarded (commutator) Green's functions 
\begin{equation}
G_{nn'}^{\textrm{R}}(t)\equiv -\frac{i}{\hbar}\:\theta(t)\langle
\textrm{GS}|[c_n^{\dagger}(t),c_{n'}]|\textrm{GS}\rangle \:,
\end{equation}
which in systems with a discrete translational symmetry depend only on $n-n'$, 
are given by the spatial Fourier transform 
\begin{equation}
G_{nn'}^{\textrm{R}}(t)=\sum_{n,n'} e^{-ik\cdot (n-n')}G^{\textrm{R}}_{-}(k,t)
\end{equation}
of the momentum-space-resolved ones [cf. Eq.~\eqref{commGF}]. They can straightforwardly 
be recast in terms of pseudospin-$1/2$ operators as~\cite{Stojanovic+:14}
\begin{equation} \label{DefCorrelFunc}
G_{nn'}^{\textrm{R}}(t)= -\frac{i}{\hbar}\:\theta(t)\langle\textrm{GS}|
[\sigma_n^{+}(t),\sigma_{n'}^{-}]|\textrm{GS}\rangle \:,
\end{equation}
where the Jordan-Wigner transformation~\cite{ColemanBOOK} [for the explicit 
form of this transformation, see Eq.~\eqref{JWtsf} above] allows one to switch 
from spinless-fermion- to the pseudospin-$1/2$ operators. It is straightforward 
to show that $G_{nn'}^{\textrm{R}}(t)$ can be rewritten in the form~\cite{Stojanovic+:14}
\begin{equation}\label{gnRt}
G_{nn'}^{\textrm{R}}(t)= \mathcal{G}^{xx}_{nn'}(t)+\mathcal{G}^{yy}_{nn'}(t)
-i[\mathcal{G}^{xy}_{nn'}(t)-\mathcal{G}^{yx}_{nn'}(t)]\:,
\end{equation}
with the retarded pseudospin-$1/2$ two-time correlation functions 
$\mathcal{G}^{\alpha\beta}_{nn'}(t)$ (\:$\alpha,\beta=x,y$\:),
which are defined as~\cite{Stojanovic+:14}
\begin{equation} \label{RetTwoTimeCorrFunc}
\mathcal{G}^{\alpha\beta}_{nn'}(t)\equiv -\frac{i}{\hbar}\:\theta(t)\langle
\textrm{GS}|[\sigma_n^{\alpha}(t),\sigma_{n'}^{\beta}]|\textrm{GS}\rangle \:.
\end{equation}
[For the sake on simplicity, the superscript $R$ has been omitted in the notation 
for these last retarded two-time correlation functions; the same convention is used
in the remainder of this section.] 

The many-body version of the Ramsey interference protocol, which can be utilized in 
all systems that are addressable at the single-qubit level, yields the real-space- and 
time-resolved commutator Green's functions of spin-$1/2$ (or pseudospin-$1/2$) operators~\cite{Knap++:13}. 
This protocol makes use a special type $R_n(\theta=\pi/2,\phi)\equiv R_n(\phi)$ of Rabi pulses, 
which are written in the general form
\begin{equation}
R_n(\theta,\phi)\equiv\mathbbm{1}_{2}\cos\frac{\theta}{2} + 
i(\sigma_n^{x}\cos\phi-\sigma_n^{y}\sin\phi)\sin\frac{\theta}{2} \:.
\end{equation}
Here $\theta=\Omega\tau$ -- with $\Omega$ being the Rabi frequency and 
$\tau$ the pulse duration -- is the pulse area and $\phi$ the phase of the 
laser field. The effect of such $\pi/2$ pulses on, e.g., the spin-down (logical-zero) 
state of a single qubit is given by $R_n(\phi)|\hspace{-0.05cm}\downarrow\rangle_n=
(|\hspace{-0.05cm}\downarrow\rangle_{n}+e^{i\phi}|\hspace{-0.05cm}\uparrow
\rangle_{n})/\sqrt{2}$.

\begin{figure}[t!]
\begin{center}
\includegraphics[width = 0.85\linewidth]{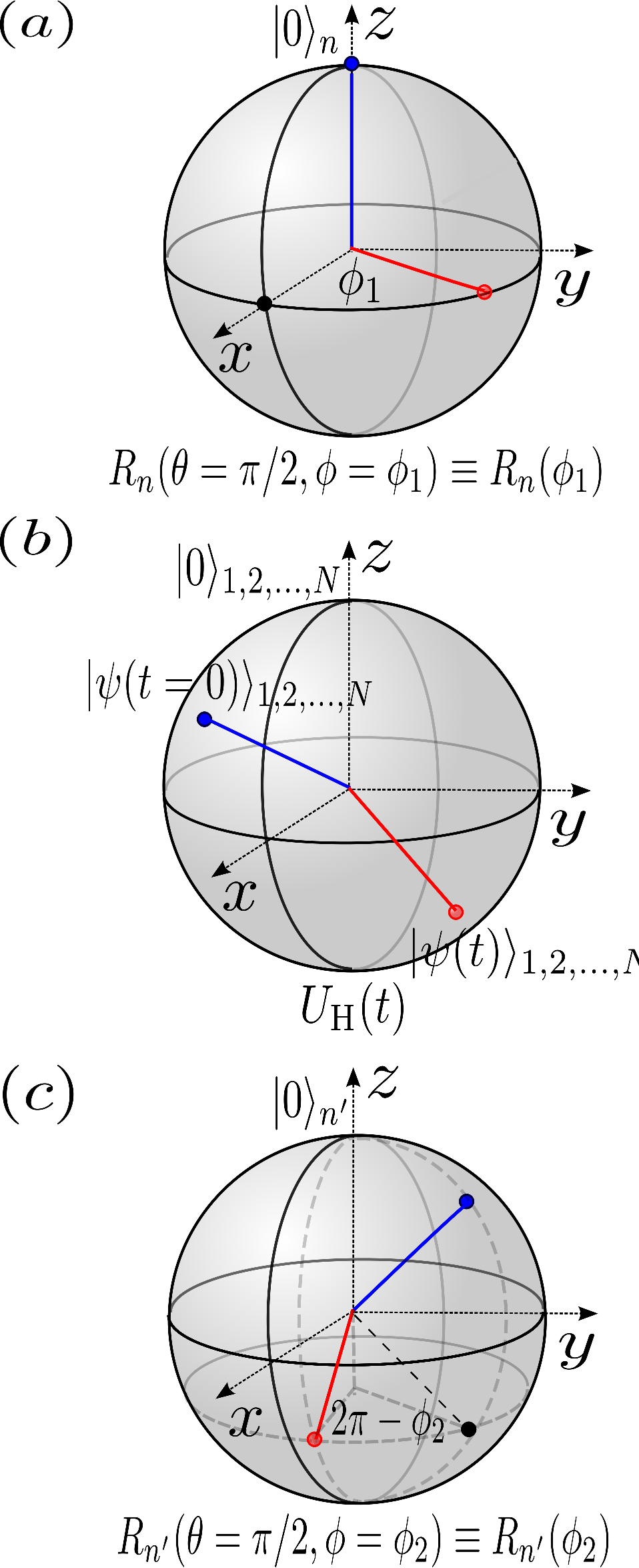}
\end{center}
\caption{\label{fig:RamseyIPillustrate}Pictorial illustration of the first
three steps of the many-body Ramsey interference protocol for measuring correlation functions $G_{nn'}
^{\textrm{R}}(t)$ [cf. Eq.~\eqref{DefCorrelFunc}]: (a) Rabi pulse $R_n(\theta=\pi/2,\phi=\phi_1)
\equiv R_n(\phi_1)$ applied to qubit $n$, which is assumed to initially be in the logical
zero state $|0\rangle_n$, illustrated on the Bloch sphere corresponding to this qubit, (b) 
free evolution of the $N$-qubit system during a time interval of duration $t$ [here $U_H(t)$ 
is the effective time-evolution operator of the system], illustrated on the Bloch sphere of
one of the qubits $1,2,\ldots,N$, and (c) Rabi pulse $R_{n'}(\theta=\pi/2,\phi=\phi_2)\equiv R_n(\phi_2)$ 
on qubit $n'$, illustrated on its corresponding Bloch sphere.}
\end{figure}

This generalized Ramsey-interference protocol consists of the following steps (for
a pictorial illustration, see Fig.~\ref{fig:RamseyIPillustrate}), akin to those used
in other applications of the Ramsey sequence~\cite{Dogra+:22}: a local $\pi/2$-rotation 
at site $n$ (with the value $\phi_1$ of the parameter $\phi$), an evolution of the 
system over the time interval of duration $t$, followed by a local $\pi/2$-rotation 
at site $n'$ or global $\pi/2$-rotation (with the value $\phi_2$ of the parameter $\phi$), 
and, finally, a measurement of $\sigma_{n'}^{z}$ (i.e. the $z$-component of the pseudospin 
at site $n'$). The result of the final measurement of $\sigma_{n'}^{z}$ is 
given by~\cite{Knap++:13}
\begin{equation} \label{MeasureResRamsey}
M_{nn'}(\phi_1,\phi_2,t)=\langle\textrm{S}_{nn'}(t)
|\:\sigma_{n'}^{z}\:|\textrm{S}_{nn'}(t)\rangle\:,
\end{equation}
where $|\textrm{S}_{nn'}(t)\rangle$ is the state obtained from $|\textrm{GS}\rangle$ 
by performing the first three steps (two $\pi/2$-rotations and a time evolution 
of duration $t$) of the Ramsey interference protocol (cf. Fig.~\ref{fig:RamseyIPillustrate}). 
In other words, the state $|\textrm{S}_{nn'}(t)\rangle$ is given by
\begin{equation}
|\textrm{S}_{nn'}(t)\rangle\equiv R_{n'}(\phi_2)U_H(t) 
R_n(\phi_1)|\textrm{GS}\rangle \:.
\end{equation}

In order to elucidate the specific form of the measurement result in Eq.~\eqref{MeasureResRamsey},
expressed in terms of the retarded correlation functions $\mathcal{G}^{\alpha\beta}_{nn'}(t)$
[cf. Eq.~\eqref{RetTwoTimeCorrFunc}], in the problem under consideration it is important to note 
that the Holstein Hamiltonian -- written in terms of pseudospin-$1/2$ (qubit) operators in 
Eq.~\eqref{H_r} -- has the $U(1)$ symmetry under pseudospin rotations around the $z$ axis and the 
one under reflections with respect to the same axis. It can straightforwardly be shown that for a 
system with these two symmetries the expression for the final measurement result in the Ramsey 
protocol [cf. Eq.~\eqref{MeasureResRamsey}] reads as follows~\cite{Stojanovic+:14}:
\begin{multline}
M_{nn'}(\phi_1,\phi_2,t)=-\frac{1}{4}\Big\{\sin(\phi_1-\phi_2)
[\:\mathcal{G}^{xx}_{nn'}(t)+\mathcal{G}^{yy}_{nn'}(t)\:] 
\\
-\cos(\phi_1-\phi_2)[\:\mathcal{G}^{xy}_{nn'}(t)
-\mathcal{G}^{yx}_{nn'}(t)\:]\Big\} \:.
\end{multline}
In particular, the terms $\mathcal{G}^{xx}_{nn'}(t)+\mathcal{G}^{yy}_{nn'}(t)$
and $\mathcal{G}^{xy}_{nn'}(t)-\mathcal{G}^{yx}_{nn'}(t)$ that one needs in order
to to determine $G_{nn'}^{\textrm{R}}(t)$ [cf. Eq.~\eqref{gnRt}] are given by 
$M_{nn'}(\phi_1,\phi_2,t)$ for $\phi_1-\phi_2=\pm \pi/2$ and $\phi_1=\phi_2$, 
respectively.

\section{Basics of the KPM} \label{KPMbasics}
In what follows, the mathematical basis for the KPM~\cite{Silver+Roeder:97} is briefly recapitulated.
The basic ideas pertaining to applications of this method for evaluating spectral functions are also 
sketched. A more detailed introduction into the KPM and its use in many-body physics can be found in
Ref.~\cite{WeisseEtAlRMP:06}.

The KPM allows one to efficiently compute dynamical properties of quantum many-body systems, 
which are typically described by spectral functions 
\begin{equation} \label{SpectralFuncObservO}
A_{O}(\omega) = \frac{1}{\pi}\:\lim_{\eta\rightarrow 0^{+}}
\textrm{Im}\:\langle \Psi_0|O^{\dagger}\:\frac{1}{\hbar\omega - H -i\eta} 
\:O|\Psi_0\rangle \:.
\end{equation}
Here $O$ is the relevant observable, while $|\Psi_0\rangle$ is the ground state of 
a system described by the Hamiltonian $H$. This quantity can be recast in the form 
\begin{equation} \label{SpectralFuncObservODelta}
A_{O}(\omega)=\sum_{s=1}^{N}|\langle\Psi_0|O|\Psi_s\rangle|^2
\:\delta(\hbar\omega+E_0 - E_s ) \:,
\end{equation}
where $E_0$ is the ground-state energy of the system and $E_s$ the energy of the
$s$-th excited state ($s=1,\ldots,N$); $|\Psi_s\rangle$ is the $s$-th energy 
eigenstate of the system. The derivation of the last expression relies on the 
completeness of the set of eigenstates $\{|\Psi_s\rangle | s=1,\ldots,N \}$
of the Hamiltonian $H$ (i.e., the relation $\sum_s |\Psi_s\rangle\langle\Psi_s|
=\mathbbm{1}$) and the use of identity in Eq.~\eqref{PlemeljFormula}. 

It is pertinent to note that the momentum-frequency resolved spectral function $A_{+}(k,\omega)$ 
[cf. Eqs.~\eqref{CpectFuncDef} and \eqref{MomFreqSpectFunc} in Sec.~\ref{genspectfunc}], is a 
special case ($O = c_k$) of the general spectral function $A_{O}(\omega)$ defined in Eq.~\eqref{SpectralFuncObservO} 
above.

The mathematical foundation of the KPM pertains to the problem of approximating a real-valued 
function $f(x)$~\cite{RivlinBOOK:81}, which is defined on the interval $[-1,1]$, by a finite 
series 
\begin{equation}\label{ChebyshevExp}
f^{(N_{\textrm{C}})}(x)=\frac{1}{\pi\sqrt{1-x^2}}\:\left[\mu^{(0)}
+2\sum_{r=1}^{N_{\textrm{C}}-1}\mu^{(r)} T_r(x) \right] \:,
\end{equation}
where $T_r(x)$ ($r=0,1,\ldots$) are Chebyshev polynomials of the first 
kind whose orthogonality relation is given by~\cite{Shima+NakayamaBOOK:10}
\begin{equation}\label{ChebyshevOrthogonal}
\int_{-1}^{1}\frac{T_r(x)T_{r'}(x)}{\pi\sqrt{1-x^{2}}}
\:dx =\frac{1+\delta_{r,0}}{2}\:\delta_{r,r'} 
\quad (\:r,r'= 0,1,\ldots\:)\:.
\end{equation}
The coefficients in the above expansion (Chebyshev moments) are given by
\begin{equation}\label{ChebyshevMoments}
\mu^{(r)}=\int_{-1}^{1} f(x)T_r(x)dx \quad 
(\:r=0,\ldots,N_{\textrm{C}}-1\:) \:.
\end{equation}
For a sufficiently smooth function $f(x)$ the last series converges uniformly to 
$f$ on any closed sub-interval of $[-1,1]$ that excludes the endpoints $\pm 1$~\cite{WeisseEtAlRMP:06}.

If the function that one aims to approximate is not continuous, the finite series in Eq.~\eqref{ChebyshevExp}
cannot converge uniformly. More precisely, this series fails to converge in the vicinity of points 
where the function $f(x)$ is not continuously differentiable. Instead, the series displays rapid Gibbs 
oscillations, the amplitude of which does not decrease in the limit where the number of terms in the series 
becomes infinite (the Gibbs phenomenon)~\cite{Shima+NakayamaBOOK:10}. However, the problem arising from the 
Gibbs phenomenon is known to be soluble for Chebyshev-polynomial expansions. In that case, for a fixed number 
$N_{\textrm{C}}$ of terms in the expansion, a set of attenuation (Gibbs damping) factors $g^{N_{\textrm{C}}}_r$ 
($r = 0,\ldots,N_{\textrm{C}}-1$) can be found, which depend implicitly on $N_{\textrm{C}}$, such that the 
modified finite-series approximants
\begin{equation}\label{ChebyshevAttenExp}
f_{\textrm{KPM}}^{(N_{\textrm{C}})}(x)=\frac{1}
{\pi\sqrt{1-x^2}}\:\Big[g^{N_{\textrm{C}}}_0\mu^{(0)}+
2\sum_{r=1}^{N_{\textrm{C}}-1}g^{N_{\textrm{C}}}_r \mu^{(r)} T_r(x) \Big]
\end{equation}
accurately reproduce the function $f(x)$ under consideration. Rephrasing, the introduction of 
these attenuation factors damps out high-frequency oscillations and constitutes the essential 
ingredient of the KPM.

The last truncation of the infinite series to order $N_{\textrm{C}}$, along with the attendant 
modification of the coefficients in the expansion [ $\mu^{(r)}\rightarrow g^{N_{\textrm{C}}}_r\mu^{(r)}$ ],
is equivalent to the convolution of the function $f(x)$ under consideration with a kernel of the 
form 
\begin{equation}\label{DefKernel}
K_{N_{\textrm{C}}}(x,y) = g^{N_{\textrm{C}}}_0 \phi_0(x)\phi_0(y) + 
2\sum_{r=1}^{N_{\textrm{C}}-1}g^{N_{\textrm{C}}}_r \phi_r(x)\phi_r(y)\:,
\end{equation}
where the function $\phi_r(x)$ is defined as 
\begin{equation}\label{DefFuncphi}
\phi_r(x)= \frac{T_r(x)}{\pi\sqrt{1-x^{2}}} \quad 
(\:r=0,\ldots,N_{\textrm{C}}-1\:) \:.
\end{equation}
In other words, one has
\begin{equation}\label{DefKernel}
f_{\textrm{KPM}}^{(N_{\textrm{C}})}(x)=\int_{-1}^{1}\:\pi
\sqrt{1-y^{2}}\: K_{N_{\textrm{C}}}(x,y)f(y)dy \:.
\end{equation}

Importantly, each choice of attenuation factors $g^{N_{\textrm{C}}}_r$ corresponds to one specific 
choice of kernel. In the problem at hand, it is pertinent to utilize the attenuation factors derived 
from the Jackson kernel~\cite{Jackson:1912}, which are given by~\cite{Silver+:96}
\begin{eqnarray} \label{JacksonAttenFactors}
g^{N_{\textrm{C}}}_r &=& \frac{1}{N_{\textrm{C}}+1}\:\Big\{\displaystyle (N_{\textrm{C}}
-r+1)\cos\left(\frac{r\pi}{N_{\textrm{C}}+1}\right) \nonumber \\
&+& \sin\left(\frac{r\pi}{N_{\textrm{C}}+1}\right)\cot\left(\frac{\pi}{N_{\textrm{C}}+1}
\right)\Big\} \:.\label{JacksonAttenFact}
\end{eqnarray}
This kernel was demonstrated to be the optimal choice for the evaluation of spectral functions 
[cf. Eqs.~\eqref{SpectralFuncObservO} and \eqref{SpectralFuncObservODelta}] using the KPM~\cite{Silver+:96}. 


\end{document}